\documentclass[12pt]{article}

\pdfoutput=1

\usepackage{ifpdf}
\ifpdf
\usepackage{graphicx,color}
\usepackage{hyperref}
\else
\usepackage[dvipdfmx]{graphicx,color}
\usepackage[dvipdfmx]{hyperref}
\fi
\usepackage{amssymb,amsfonts,amsmath,cancel,cite,multirow}
\usepackage[capitalise]{cleveref}
\usepackage{slashed}

\newcommand{\eqs}[1]{\begin{equation}\begin{split} #1 \end{split}\end{equation}}

\begin{document}

\begin{titlepage}

\begin{flushright}
	CTPU-PTC-19-25
\end{flushright}

\vskip 1.35cm
\begin{center}

{\large
\textbf{
Lessons from $T^{\mu}_{~ \mu}$ on inflation models: \\
two-loop renormalization of $\eta$ in the scalar QED}}
\vskip 1.2cm

Ayuki Kamada$^{a}$
and Takumi Kuwahara$^{a}$

\vskip 0.4cm

\textit{$^a$
Center for Theoretical Physics of the Universe,
Institute for Basic Science (IBS), Daejeon 34126, Korea
}

\vskip 1.5cm

\begin{abstract}
A non-minimal coupling $\eta$ has been attracting growing interest particularly in the context of inflation models, though its quantum nature is not clear yet.
We study the renormalization of a non-minimal coupling in the scalar quantum electrodynamics (QED).
We find {\it no} inhomogeneous term of the renormalization group equation (RGE) at the two-loop level.
This is similar to other theories, where an inhomogeneous term of the RGE appears only at a higher-loop order: e.g., four-loop order in $\lambda \phi^{4}$ theory.
\end{abstract}

\end{center}
\end{titlepage}

\section{Introduction}

Properties of the energy-momentum tensor $T_{\mu \nu}$ are of particular interest in quantum field theory even in the flat spacetime.
It determines the linear response of matter to external gravitons.
It is a single place where we can study a non-minimal coupling of a scalar field $\phi$ to gravity, $\xi R \phi^{2}$ ($R$: Ricci scalar), in the flat spacetime.
It provides an improvement term, $- (\partial_{\mu} \partial_{\nu} - g_{\mu \nu} \partial^{2}) \phi^{2}$, to $T_{\mu \nu}$ in the flat spacetime.
The improvement term does not change the conservation of $T_{\mu \nu}$ nor the corresponding Poincar\'e algebra~\cite{Callan:1970ze, Coleman:1970je}.

The improvement term, on the other hand, changes the trace of the energy-momentum tensor $T^{\mu}_{~ \mu}$ by $2 \eta \partial^{2} \phi^{2}$.
Here $\xi = \xi_{c} + \eta / (d - 1)$ with the conformal coupling being $\xi_{c} = (d - 2) / (4 (d - 1))$ in $d$ dimensions.
The improvement term leads to an uncertainty of $T_{\mu \nu}$: $T_{\mu \nu}$ that couples to gravity does not need to satisfy $T^{\mu}_{~ \mu} = 0$ even if matter is at the conformal fixed point.
Meanwhile, the improvement term is subject to renormalization, although $T_{\mu \nu}$ is na\"ively expected to be finite due to its conservation.
Namely, one has to renormalize $\eta$.

The renormalization of $\eta$ is of interest as it is and studied intensively in $\lambda \phi^{4}$ theory~\cite{Collins:1976vm, Brown:1980qq, Hathrell:1981zb}.
The key is that the renormalization group equation (RGE) of $\eta$ is inhomogeneous.
The inhomogeneous term arises from the renormalization of trace-anomaly terms (i.e., composite operators~\cite{Zimmermann:1969jj, Lowenstein:1971jk, Collins:1974da, Breitenlohner:1977hr}).
The trace anomaly is associated with violation of conformal symmetry at a quantum level and provides additional terms to $T^{\mu}_{~ \mu}$~\cite{Callan:1970ze, Coleman:1970je, Freedman:1974gs, Freedman:1974ze, Collins:1976vm, Nielsen:1977sy, Adler:1976zt, Collins:1976yq, Brown:1979pq, Brown:1980qq, Hathrell:1981zb, Hathrell:1981gz}.

The inhomogeneous term of the RGE induces the inhomogeneous solution of $\Delta \eta$ irrespectively of our initial choice of $\eta$ and thus $\Delta \eta$ is a quantum-induced value of $\eta$~\cite{Kamada:2019euz}.
The inhomogeneous solution of the RGE would lead to an important consequence in phenomenology.
For example, a small change in the non-minimal coupling would lead to a detectable change in the prediction of the inflation dynamics: the chaotic inflation with a simple power-law potential~\cite{Linde:1983gd}, whose large tensor-to-scalar ratio is disfavored by measurements of cosmic microwave background (CMB) anisotropies~\cite{Hinshaw:2012aka, Akrami:2018odb}, gets viable again just with $\xi \sim - 10^{-3}$~\cite{Linde:2011nh, Boubekeur:2015xza, Shokri:2019rfi}.%
\footnote{
The minus sign originates from our convention following Refs.~\cite{Kolb:1990vq, Hathrell:1981zb}: the metric signature is $(+ , - , - , -)$; the Einstein-Hilbert action with a free singlet scalar is
\begin{eqnarray}
S_{\text{E-H}} = - \frac{M_{\rm pl}^{2}}{2} \int d^{4} x \sqrt{- g} R + \int d^{4} x \sqrt{- g} \left( \frac{1}{2} g^{\mu \nu} \nabla_{\mu} \phi \nabla_{\nu} \phi + \frac{1}{2} \xi R \phi^{2} \right)  \,,
\end{eqnarray}
with the reduced Planck mass $M_{\rm pl}$;
and the four-dimensional conformal coupling is $\xi_{c} = + 1/6$.
}

Meanwhile, an inhomogeneous term is expected to appear first at a higher-loop order: four-loop order in $\lambda \phi^{4}$ theory~\cite{Collins:1976vm, Brown:1980qq, Hathrell:1981zb}; and at least two-loop order in the two scalar theory and Yukawa theory~\cite{Kamada:2019euz}.
We devote this article to investigating the RGE of $\eta$ and its inhomogeneous term in the scalar quantum electrodynamics (QED) at the two-loop level.
We also note how our results are expected to be generalized to the scalar quantum chromodynamics (QCD).
For the fermion QED and QCD, the trace anomaly is known at all orders~\cite{Nielsen:1977sy, Adler:1976zt, Collins:1976yq, Hathrell:1981gz}.
We find {\it no} inhomogeneous term at the two-loop order.

This article is organized as follows.
In the next section, we discuss a general property of the RGE of $\eta$ to identify what to be computed.
We show our computation results at the one-loop and two-loop orders in \cref{sec:oneloop} and \cref{sec:twoloop}, respectively.
\cref{sec:discussion} is devoted to discussing the implications of the result.
Throughout this article, we adopt the modified minimal subtraction ($\overline{\rm MS}$) scheme~\cite{Ashmore:1972uj, Bollini:1972ui, tHooft:1972tcz} with the spacetime dimensions of $d = 4 - \epsilon$ and the (modified) renormalization scale $\mu$ ($\tilde \mu$).

\section{RGE of $\eta$ \label{sec:RGE-eta}}
The scalar-QED action is
\eqs{
S_{\rm mat} =& \int d^{4} x \sqrt{- g} \left( - \frac{1}{4} g^{\mu \lambda} g^{\nu \kappa} F_{0 \mu \nu} F_{0 \lambda \kappa} +  g^{\mu \nu} D_{0 \mu} \phi_{0}^{*} D_{0 \nu} \phi_{0} + \xi_{0} R |\phi_{0}|^{2} - m_{0}^{2} |\phi_{0}|^{2} - \frac{1}{4} \lambda_{0} |\phi_{0}|^{4} \right) \\
& + S_{\rm fix} \,,
}
with $D_{\mu}$ being the gauge and diffeomorphism covariant derivative.
We omit the gauge fixing term $S_{\rm fix}$ except for the gauge boson propagator as discussed in Ref.~\cite{Kamada:2019pmx}.
Parameters are a gauge coupling $e_{0}$ (charge being $Q$), a scalar mass $m_{0}$, and a quartic coupling $\lambda_{0}$.
The subscripts $0$ denote the bare fields and couplings.
We provide details of multiplicative renormalization in \cref{sec:scalarQED}.
We remark that in the flat spacetime, the non-minimal coupling $\xi_{0} = \xi_{c} + \eta_{0} / (d - 1)$ does not affect multiplicative renormalization.

We define the energy-momentum tensor as a linear response of the matter action to the metric:
\eqs{
\label{eq:Tmunu}
T^{\mu \nu} = - \frac{2}{\sqrt{-g}} \frac{\delta S_{\rm mat} \left[ \{ \phi_{0 i} \}, g_{\mu \nu}; \{ \lambda_{0 a} \} \right]}{\delta g_{\mu \nu}} \,.
}
The $d$-dimensional flat-spacetime energy-momentum tensor is given by
\eqs{
T_{\mu \nu} =& - g^{\lambda \kappa} F_{0 \mu \lambda} F_{0 \nu \kappa} + 2 D_{0 \mu} \phi_{0}^{*} D_{0 \nu} \phi_{0} - 2 \left( \xi_{c} + \frac{\eta_{0}}{d - 1} \right) (\partial_{\mu} \partial_{\nu} - g_{\mu \nu} \partial^{2}) |\phi_{0}|^{2} \\
& - g_{\mu \nu} \left( - \frac{1}{4} F_{0 \lambda \kappa}^{2} + |D_{0 \mu} \phi_{0}|^{2} - m_{0}^{2} |\phi_{0}|^{2} - \frac{1}{4} \lambda_{0} |\phi_{0}|^{4} \right) \,.
}
Taking the trace, one finds
\eqs{
\label{eq:scalarclassicalT}
T^{\mu}_{~ \mu} = \epsilon \left( - \frac{1}{4} F_{0 \mu \nu}^{2} + \frac{1}{4} \lambda_{0} |\phi_{0}|^{4} \right) + 2 m_{0}^{2} |\phi_{0}|^{2} + 2 \eta_{0} \partial^{2} |\phi_{0}|^{2} + ({\rm e.o.m.}) \,,
}
where the last term is proportional to the equation of motion as
\eqs{
({\rm e.o.m.})  =& \left(1 - \frac{\epsilon}{2} \right) \phi_{0}^{*} \left( D_{0}^{2} \phi_{0} + m_{0}^{2} \phi_{0} + \frac{2}{4} \lambda_{0} |\phi_{0}|^{2} \phi_{0} \right) \\
& + \left(1 - \frac{\epsilon}{2} \right) \left( D_{0}^{2} \phi_{0}^{*} + m_{0}^{2} \phi_{0}^{*} + \frac{2}{4} \lambda_{0} |\phi_{0}|^{2} \phi_{0}^{*} \right) \phi_{0} \,.
}

In the flat-spacetime, $T^{\mu}_{~ \mu}$ is almost pre-determined by the multiplicative renormalization of the fields and parameters.
The single exception is the non-minimal coupling $\eta_{0}$, whose renormalization is determined by the renormalization of $T^{\mu}_{~ \mu}$ itself.
We renormalize $\eta_{0}$ as
\eqs{
Z_{\phi} \eta_{0} = Z_{\eta} \eta \,,
}
with the wavefunction renormalization of $\phi$ being $Z_{\phi}^{1/2}$.
As discussed in Ref.~\cite{Kamada:2019euz}, the RGE of $\eta$ takes a form of
\eqs{
\label{eq:beta-eta}
\frac{d \eta}{d \ln \mu} = \gamma_{\phi^{2}}^{T} \eta + {\tilde \beta}_{\eta} \,,
}
where $\eta$ and $\gamma_{\phi^{2}}$ should be understood as a vector and matrix, respectively, for multiple scalar fields.
Here
\eqs{
\label{eq:gamma}
& \phi_{0}^{2} = Z_{\phi^{2}} [\phi^{2}]  \,, \\
& \frac{d \ln Z_{\phi^{2}}}{d \ln \mu} = \gamma_{\phi^{2}} \,,
}
and a square bracket denotes the renormalized composite operator.
The homogeneous term is proportional to the anomalous dimension of $\phi^{2}$ in the RGE.
This is because the renormalization of the scalar field squared is multiplicative, $Z_{\phi^{2}}^{-1}  \partial^{2} \phi_{0}^{2} = \partial^{2} [\phi^{2}]$.
It means that all the counterterms to renormalize $\phi^{2}$ is included in $Z_{\phi^{2}}^{-1} Z_{\rm \phi}$.
${\tilde \beta}_{\eta}$ denotes the inhomogeneous term of the RGE and induces $\Delta \eta$ through the running irrespectively of our initial choice of $\eta$.
This $\Delta \eta$ is nothing but a quantum-induced value of $\eta$.

In the following, we determine ${\tilde \beta}_{\eta}$ in the scalar QED by determining $Z_{\eta}$ at the two-loop level.
If ${\tilde \beta}_{\eta}$ appears at the 2-loop level, $\Delta \eta$ is the 1-loop order, since $\gamma_{\phi^{2}}$ appears at the one-loop level.
We take the Feynman-'t Hooft gauge, but note that $\eta$ is a gauge-invariant quantity and thus its RGE does not depend on the gauge choice.
For diagrammatic convenience, we introduce the ``scalaron'' $\sigma$ that couples to $T^\mu_{~\mu}$ as
\eqs{
\mathcal{L}_{\sigma\text{-mat}} = \sigma T^\mu_{~\mu} \,.
}
We calculate the amputated amplitude of scalaron decay into light scalars: $\sigma (p) \to \phi^{*} (q) \phi (k)$, where $p$, $q$, and $k$ are external momenta.
Although we consider scalaron decay, our analysis is applicable even to studying the properties of non-minimal couplings in the models without the scalaron.

The leading contributions originate from 
\eqs{
T^{\mu}_{~ \mu} \supset - \frac{1}{4} \epsilon F^{2}_{\mu \nu} + 2 m^{2} |\phi|^{2} + 2 \eta \partial^{2} |\phi|^{2} - \frac{1}{4} \epsilon (Z_{A} - 1) F^{2}_{\mu \nu} + 2 (Z_{m^{2}} - 1) m^{2} |\phi|^{2} + 2 (Z_{\eta} - 1) \eta \partial^{2} |\phi|^{2} \,,
\label{eq:emtensor_sqed}
}
where we use the renormalized fields and couplings.
Hereafter we assume that the quartic coupling $\lambda$ is negligible.
As stressed in Ref.~\cite{Kamada:2019pmx}, a key point is that $T^{\mu}_{~ \mu}$ contains terms proportional to $\epsilon$: $T^{\mu}_{~ \mu} \supset - (1/4) \epsilon F^{2}_{0 \mu \nu}$.
These terms vanish in the limit of $\epsilon \to 0$ at the classical level, but not at the quantum level due to the renormalization of composite operators: $F^{2}_{0 \mu \nu}$.
This is the origin of the trace anomaly.


\section{One-loop order \label{sec:oneloop}}

\begin{figure}
	\centering
	\includegraphics[width=0.7\linewidth]{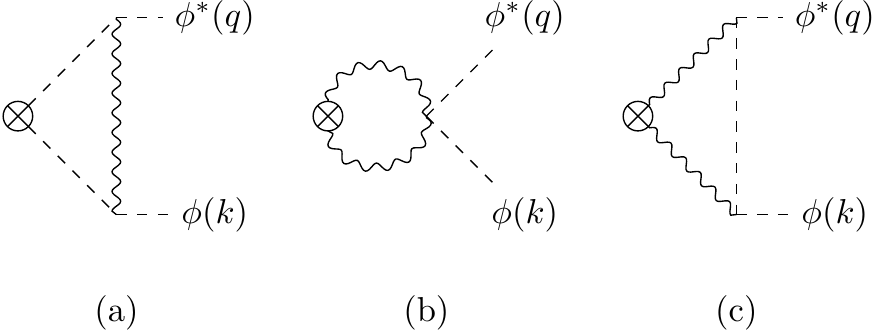}
	\caption{One-loop diagrams for scalaron decay $\sigma (p) \to \phi^{*} (q) \phi (k)$.
	Crossed dots denote insertion of the energy-momentum tensor, $2 m^{2} |\phi|^{2} + 2 \eta \partial^{2} |\phi|^{2}$ (a), and $- (1/4) \epsilon F^{2}_{\mu \nu}$ (b, c).}
	\label{fig:eta}
\end{figure}

\cref{fig:eta} shows one-loop diagrams contributing to $\sigma (p) \to \phi^{*} (q) \phi (k)$ in the scalar QED.
We summarize one-loop calculations in \cref{sec:oneloop_app}.
\cref{fig:eta} (a) from the insertion of $T^{\mu}_{~ \mu} \supset 2 m^{2} |\phi|^{2} + 2 \eta \partial^{2} |\phi|^{2}$.
The divergent part [see \cref{eq:etaa} for the full expression] is
\eqs{
\left( i {\cal M}_{(a)}^{\rm loop} \right)^{\rm pole}_{\rm of \, \epsilon} = 4 i (m^{2} - \eta p^{2}) \frac{Q^{2} e^{2}}{16\pi^{2}} \frac{1}{\epsilon} \,.
}
\cref{fig:eta} (b) and (c) originate from the insertion of $T^{\mu}_{~ \mu} \supset - (1/4) \epsilon F^{2}_{\mu \nu}$ and only provide finite contributions:
\eqs{
\label{eq:M_loop_b_c}
i {\cal M}_{(b)}^{\rm loop} = 3 i \frac{Q^{2} e^{2}}{16 \pi^{2}} p^{2} \,, \quad 
i {\cal M}_{(c)}^{\rm loop} = - 6 i \frac{Q^{2} e^{2}}{16 \pi^{2}} (k \cdot q) \,.
}
By summing them up, we obtain
\eqs{
\label{eq:M_loop_bc}
i {\cal M}_{(b) + (c)}^{\rm loop} = 6 i \frac{Q^{2} e^{2}}{16 \pi^{2}} m^{2} \,.
}
This contribution is reproduced by the insertion of $- \beta_{m^{2}} |\phi|^{2}$, where $\beta_{m^{2}}$ is the $\beta$ function of $m^{2}$ in \cref{eq:betam}.
Meanwhile, there is no contribution proportional to $p^{2}$, which can be regarded as a finite value of $\eta$.

The divergence of $i {\cal M}_{(a)}^{\rm loop}$ is canceled by the counterterm contributions of $T^{\mu}_{~ \mu} \supset 2 (Z_{m^{2}} - 1) m^{2} |\phi|^{2} + 2 (Z_{\eta} - 1) \eta \partial^{2} |\phi|^{2}$.
We again note that $Z_{m^{2}} - 1$ is pre-determined by the self-energy of $\phi$ [see \cref{eq:scalarQED-Z}].
The counterterm of the non-minimal coupling is determined to absorb this divergence as
\eqs{
Z_{\eta} - 1 = - 2 \frac{Q^{2} e^{2}}{16 \pi^{2}} \frac{1}{\epsilon} \,.
}
From $Z_{\phi}$ in \cref{eq:scalarQED-Z}, one obtains
\eqs{
\frac{d \eta}{d \ln \mu} = - 6 \frac{Q^{2} e^{2}}{16 \pi^{2}} \eta \,,
}
at the one-loop level.
At this order, we only find an homogeneous term of the RGE and we also confirm the expression of \cref{eq:beta-eta} from \cref{eq:scalarQED-Zsqed}.
The solution is
\eqs{
\label{eq:scalarQED-solution}
\eta = \eta_{i} \left( \frac{e^{2}}{e_{i}^{2}} \right)^{- 9} \,.
}
Here, the subscript $i$ denotes the boundary condition for the RGE: $\eta = \eta_i$ at $e = e_i$.

We note that the above discussion does not change for the scalar QCD.
In the scalar QCD, $Q^{2}$ is replaced by $T(S)$ denoting the one-half of the Dynkin index of the representation for scalar fields. 
The analytic solution \cref{eq:scalarQED-solution} depends on the beta function of $e$ at the one-loop level, and thus the power of the coupling $e$ in \cref{eq:scalarQED-solution} will change in the scalar QCD.

\section{Two-loop order \label{sec:twoloop}}

We focus only on the renormalization of $\eta$, which is determined by the divergent part of the diagrams.
Contributions to the non-minimal coupling are proportional to an incoming momentum squared $p^{2}$.
Thus, we take the massless limit of $\phi$: $m \to 0$.
We also focus on the inhomogeneous term of the RGE ${\tilde \beta}_{\eta}$, which originates from the insertion of $T^{\mu}_{~ \mu} \supset - (1/4) \epsilon F^{2}_{0 \mu \nu}$.

There are two types of two-loop contributions from the insertion of $T^{\mu}_{~ \mu} \supset - (1/4) \epsilon F^{2}_{0 \mu \nu}$: 1) one-loop diagrams with the insertion of $T^{\mu}_{~ \mu} \supset (\beta_{e} / 2e) F^{2}_{\mu \nu}$; and 2) two-loop-order diagrams with the insertion of $T^{\mu}_{~ \mu} \supset - (1/4) \epsilon F^{2}_{\mu \nu}$.
Here $\beta_{e}$ is the one-loop $\beta$ function of $e$ [see \cref{eq:scalarQED-betae}], which is obtained from $\epsilon (Z_{A} - 1) = - 2 \beta_{e} / e$ at the one-loop level.
We remark that this simple relation between $\beta_{e}$ and $Z_{A}$ holds only in the QED because of the Ward-Takahashi identity.
In the QCD, to obtain the $\beta$ function from $T^{\mu}_{~ \mu} \supset - (1/4) \epsilon F^{2}_{0 \mu \nu}$, one has to take into account self-couplings of the gauge boson.
We demonstrate it in \cref{sec:YM}, since intriguingly it gives another derivation of the one-loop $\beta$ function in the QCD.

First, we consider the type-1) contributions.
They are given by \cref{fig:eta} (b) and (c) with the insertion of $T^{\mu}_{~ \mu} \supset (\beta_{e} / 2e) F^{2}_{\mu \nu}$ instead of $T^{\mu}_{~ \mu} \supset - (1/4) \epsilon F^{2}_{\mu \nu}$.
The divergent parts are
\eqs{
\left( i {\cal M}_{(b)}^{\text{1-loop}} \right)^{\rm pole}_{\rm of \, \epsilon} = 
- \frac{2 \beta_{e}}{e} \frac{1}{\epsilon} i {\cal M}_{(b)}^{\text{loop}} \,, \quad
\left( i {\cal M}_{(c)}^{\text{1-loop}} \right)^{\rm pole}_{\rm of \, \epsilon} = 
- \frac{2 \beta_{e}}{e} \frac{1}{\epsilon} i {\cal M}_{(c)}^{\text{loop}} \,.
}
where $i {\cal M}^{\rm loop}_{(b)}$ and $i {\cal M}^{\rm loop}_{(c)}$ are shown in \cref{fig:eta} and given by \cref{eq:M_loop_b_c}.
As before [\cref{eq:M_loop_bc}], the terms proportional to $p^{2}$ cancel with each other.
Thus, there is no contribution to ${\tilde \beta}_{\eta}$.

\begin{figure}
	\centering
	\includegraphics[width=1.0\linewidth]{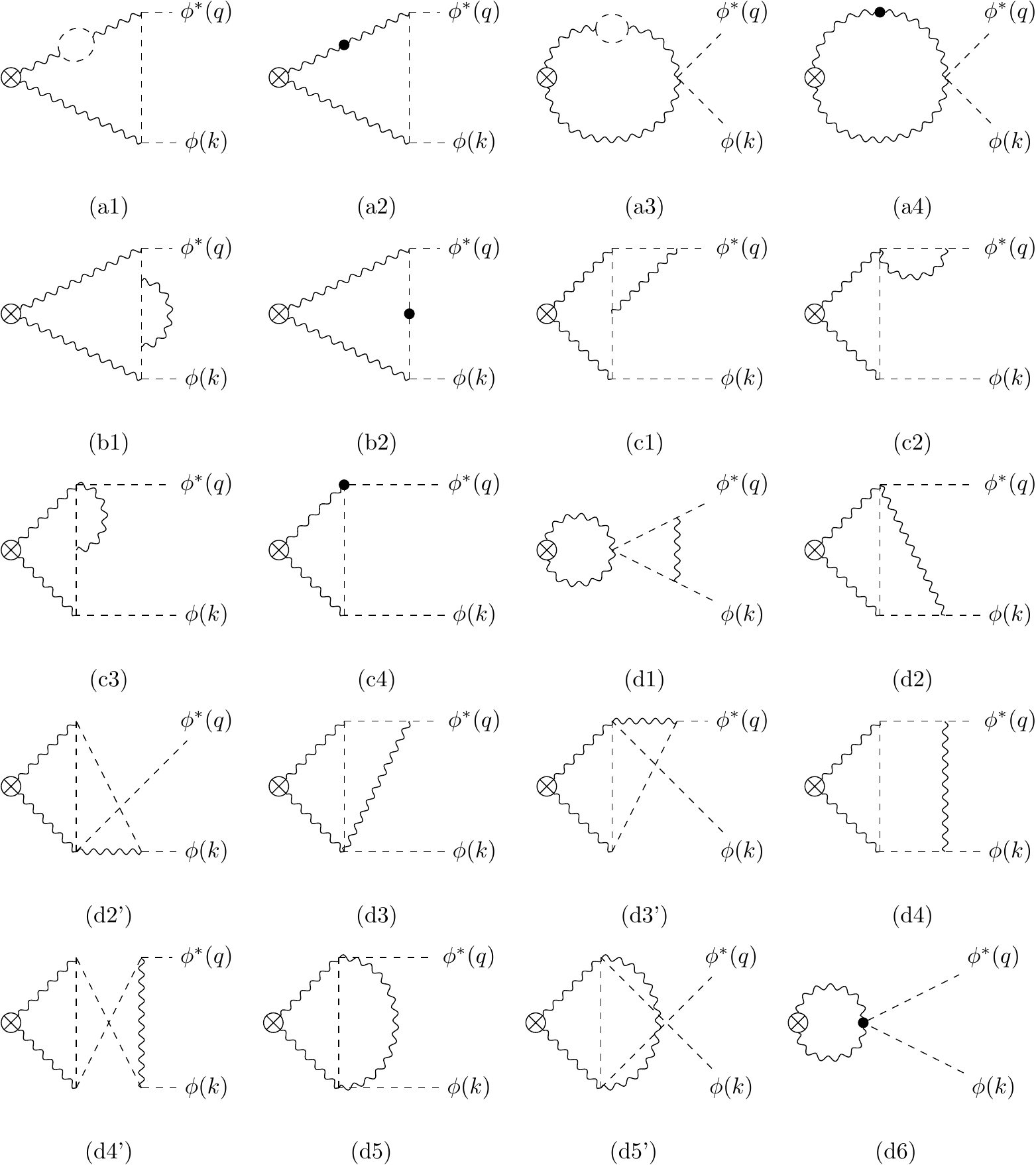}
	\caption{Two-loop-order diagrams for scalaron decay $\sigma (p) \to \phi^{*} (q) \phi (k)$.
	For (a1)-(a4) and (c1)-(c4), there are contributions also from diagrams with internal photons being exchanged.
	For (a1) and (a3), there are also contributions from the $A_{\mu}^{2} |\phi|^{2}$ vertex, but they vanish in the massless limit of $m \to 0$ in the dimensional regularization.
	Crossed dots denote the insertion of the energy-momentum tensor, $- (1/4) \epsilon F^{2}_{\mu \nu}$.
	Filled dots denote the insertion of the counterterms, $- (1/4) (Z_{A} - 1) F^{2}_{\mu \nu}$ (a2, a4), $(Z_{\phi} - 1) |\partial_{\mu} \phi|^{2}$ (b2), $i Q (Z_{e} - 1) {\tilde \mu}^{\epsilon / 2} e A^{\mu} (\phi^{*} \partial_{\mu} \phi - \partial_{\mu} \phi^{*} \phi)$ (c4), and $Q^{2} (Z_{e^{2}} - 1) {\tilde \mu}^{\epsilon} e^{2} A_{\mu}^{2} |\phi|^{2}$ (d6).}
	\label{fig:twoloop}
\end{figure}

Next, we consider the type-2) contributions.
They are given by \cref{fig:twoloop} with the insertion of $T^{\mu}_{~ \mu} \supset - (1/4) \epsilon F^{2}_{\mu \nu}$.
For (a1)-(a4) and (c1)-(c4), there are contributions also from diagrams with internal photons being exchanged.
For (a1) and (a3), there are also contributions from the $A_{\mu}^{2} |\phi|^{2}$ vertex, but they vanish in the massless limit of $m \to 0$ in the dimensional regularization.
There are two types of diagrams among the type-2) contributions: two-loop diagrams {\it without} the insertion of counterterms and one-loop diagrams {\it with} the insertion of one-loop counterterms (black dots in \cref{fig:twoloop}).
Though direct evaluation of two-loop diagrams is beyond the scope of this paper, we evaluate the $1 / \epsilon$ pole [i.e., $(1 / \epsilon)^{2}$ pole except for $\epsilon$ from $- (1/4) \epsilon F^{2}_{\mu \nu}$] based on the {\it finiteness} of renormalizable theory like the scalar QED (concretely, renormalizability of the composite operator, $T^{\mu}_{~ \mu} \supset F^{2}_{\mu \nu}$) in light of the BPHZ theorem~\cite{Bogoliubov:1957gp, Hepp:1966eg, Zimmermann:1969jj}.
In the dimensional regularization, two-loop diagrams give the leading divergence of
\eqs{
\left. i {\cal M}^{\text{2-loop}} \right.|^{\rm leading}_{\rm div} = i \epsilon 
p^{2} \left( \frac{Q^{2} e^{2}}{16 \pi^{2}} \frac{1}{\epsilon} \right)^{2} {\mu}^{2 \epsilon} F^{\text{2-loop}} (p, q, k) \,,
}
where $F^{\text{2-loop}} (p, q, k)$ is finite.
Meanwhile, one-loop diagrams {\it with} the insertion of counterterms give
\eqs{
\left. i {\cal M}^{\text{1-loop}} \right.|_{\rm div} = i \epsilon 
p^{2} \left( \frac{Q^{2} e^{2}}{16 \pi^{2}} \frac{1}{\epsilon} \right) \left( \frac{Q^{2} e^{2}}{16 \pi^{2}} \frac{1}{\epsilon} \right) \mu^{\epsilon} F^{\text{1-loop}} (p, q, k) \,,
}
where $F^{\text{1-loop}} (p, q, k)$ is finite.
The former $(Q^{2} e^{2} / 16 \pi^{2}) (1 / \epsilon)$ comes from the one-loop counterterm and the latter $(Q^{2} e^{2} / 16 \pi^{2}) (1/ \epsilon) \mu^{\epsilon}$ comes from the one-loop diagram.
The finiteness of the theory means that there is no non-local divergences and thus no $\epsilon (1 / \epsilon) \ln {\mu}$, relating numerical coefficients of the above two contributions: 
\eqs{
2 F^{\text{2-loop}} + F^{\text{1-loop}} = 0 \,.
}
Thus, we only need to evaluate one-loop diagrams to obtain $F^{\text{2-loop}}$.

In \cref{sec:twoloop_app}, we take a closer look at two-loop-order diagrams and their subdivergences.
The one-loop diagrams with the insertion of counterterms give
\eqs{
& i {\cal M}^{\text{1-loop}}_{(a2)} = - 2 (Z_{A} - 1) i {\cal M}^{\rm loop}_{(c)} \,, \quad i {\cal M}^{\text{1-loop}}_{(a4)} = - 2 (Z_{A} - 1) i {\cal M}^{\rm loop}_{(b)} \,, \\
& i {\cal M}^{\rm 1-loop}_{(b2)} = - (Z_{\phi} - 1) i {\cal M}^{\rm loop}_{(c)} \,, \quad i {\cal M}^{\text{1-loop}}_{(c4)} = 2 (Z_{e} - 1) i {\cal M}^{\rm loop}_{(c)} \,, \\
& i {\cal M}^{\text{1-loop}}_{(d6)} = (Z_{e^{2}} - 1) i {\cal M}^{\rm loop}_{(b)} \,,
}
where $i {\cal M}^{\rm loop}_{(b)}$ and $i {\cal M}^{\rm loop}_{(c)}$ are shown in \cref{fig:eta} and given by \cref{eq:M_loop_b_c}.
Noting the Ward-Takahashi identity, $Z_{\phi} = Z_{e} = Z_{e^{2}}$, and also that $i {\cal M}^{\rm loop}_{(b) + (c)}$ does not give any term proportional to $p^{2}$, we find that $F^{\text{1-loop}} = 0$ and thus $F^{\text{2-loop}} = 0$.
In summary, there is no contribution to ${\tilde \beta}_{\eta}$.

The above discussion is not directly applicable to the massless scalar QCD.
There are more diagrams appearing from the self-couplings of the gauge boson.
On the other hand, we would expect that there is no contribution to ${\tilde \beta}_{\eta}$ at the two-loop level even in the massless scalar QCD.
We discuss the physical reason in the next section.
Though it is beyond the scope of this paper, it will be important to note the followings to see it in a similar calculation to this section: the Slavnov-Taylor identity~\cite{Taylor:1971ff, Slavnov:1972fg} gives $(Z_{e^{2}} - 1) = 2 (Z_{e}^{2} - 1) - (Z_{\phi} - 1)$ at the one-loop level; and $i \mathcal{M}^{\rm loop}_{(b) + (c)}$ does not provide the $p^{2}$-term at the one-loop level as seen in the previous section.


\section{Discussion \label{sec:discussion}}

The energy-momentum tensor is the single (and thus valuable) place where we can study the properties of a non-minimal coupling in the flat spacetime.
We have studied the RGE of $\eta$ in the scalar QED at the two-loop level, through the renormalization of $T^{\mu}_{~ \mu}$.

We have found an homogeneous term of the RGE at the one-loop order.
In the scalar QED, the homogeneous solution of the RGE blows up toward a low energy as the gauge theory becomes more weakly coupled.
This is because of $\gamma_{\phi^{2}} < 0$, in contrast to the $\lambda \phi^{4}$ theory~\cite{Collins:1976vm, Brown:1980qq, Hathrell:1981zb} and Yukawa theory~\cite{Kamada:2019euz}.
Since the scalar mass squared follows $\beta_{m^{2}} =  \gamma_{\phi^{2}} m^{2}$, the scalar mass squared also blows up toward a low energy.
It means that the scalar QED does not approach a massless (conformal) free scalar theory at a low energy, without fine-tuning.

On the other hand, we have found {\it no} inhomogeneous 
term of the RGE even at the two-loop level, i.e., no inhomogeneous solution of the RGE at the one-loop order, as in the $\lambda \phi^{4}$ theory~\cite{Collins:1976vm, Brown:1980qq, Hathrell:1981zb}.
We may expect this result in a generic class of theories for the following reasoning.
For the usual quantities without any reason, we expect the inhomogeneous solution of the RGE at the one-loop order.
For example, in scalar QCD, the quartic coupling (which we assume zero) is induced by the gauge coupling at the one-loop order (though it does not change the above discussion).
In contrast, we may expect the inhomogeneous solution of the non-minimal coupling $\eta$ at the two-loop order or higher. 
The inhomogeneous solution of $\eta$ is expected to originate from the quantum breaking of scale invariance, i.e., the trace anomaly which appears only at the loop order.
It means that an extra loop order or higher is required to induce $\eta$ only from the usual couplings (such as gauge, Yukawa, and quartic couplings).
We will check this expectation, i.e., {\it no} inhomogeneous term of the RGE at the two-loop level, in the two-scalar theory, Yukawa theory, and scalar QCD, somewhere else.

Finally we remark that there is one important subtlety when one identifies $\phi$ as the inflaton.
During inflation, $\phi$ may take a field value, which breaks the gauge invariance and gives mass to gauge bosons.
In this case, an additional contribution to the threshold correction may appear from diagrams with the gauge boson mass term being inserted.
We will also study it somewhere else.

\subsection*{Acknowledgement}
The work of A. K. and T. K. is supported by IBS under the project code, IBS-R018-D1.
A. K. thanks Kazuya Yonekura for valuable advices.

\bibliographystyle{./utphys}
\bibliography{./ref}

\newpage
\appendix

\section{Scalar QED \label{sec:scalarQED}}
We use the $\overline{\rm MS}$ scheme with the spacetime dimensions of $d = 4 - \epsilon$ and the renormalization scale $\mu$.
We compensate a mass dimension by the modified renormalization scale ${\tilde \mu}$, which is defined as
\eqs{
{\tilde \mu}^{2} = \mu^{2} \frac{e^{\gamma_{E}}}{4 \pi}
}
with $\gamma_{E} \simeq 0.577$ being Euler's constant.
We use Feynman-'t Hooft gauge $(\xi^\mathrm{gf} = 1)$ in the loop calculations.

The Lagrangian density is%
\eqs{
\label{eq:scalarQEDL}
{\cal L} = - \frac{1}{4} F_{0 \mu \nu}^{2} - \frac{1}{2 \xi_{0}^{\rm gf}} (\partial_{\mu} A_{0}^{\mu})^{2} + |D_{0 \mu} \phi_{0}|^{2} - m_{0}^{2} |\phi_{0}|^{2} - \frac{1}{4} \lambda_{0} |\phi_{0}|^{4} \,,
}
with $D_{0 \mu} = \partial_{\mu} - i Q e_{0} A_{0 \mu}$ being the gauge covariant derivative for a charge $Q$.
We have integrated out the Nakanishi-Lautrup field and (anti-)ghost fields.

The multiplicative renormalization is set for fields as
\eqs{
\phi_{0} = Z_{\phi}^{1/2} \phi \,, \quad A_{0 \mu} = Z_{A}^{1/2} A_{\mu}
}
and for parameters as
\eqs{
& Z_{\phi} Z_{A}^{1/2} e_{0} = Z_{e} {\tilde \mu}^{\epsilon / 2} e \,, \quad Z_{\phi} Z_{A} e_{0}^{2} = Z_{e^{2}} {\tilde \mu}^{\epsilon} e^{2} \quad ({\rm i.e.}, Z_{\phi} Z_{e^{2}}= Z_{e}^{2}) \,, \\
& Z_{\phi} m_{0}^{2} = Z_{m^{2}} m^{2} \,, \quad Z_{\phi}^{2} \lambda_{0} = Z_{\lambda} {\tilde \mu}^{\epsilon} \lambda \,, \quad Z_{A} / \xi^{\rm gf}_{0} = Z_{\xi^{\rm gf}} / \xi^{\rm gf} \,,
}
and
\eqs{
Z_{\phi} \eta_{0} = Z_{\eta} \eta \,.
}
The Lagrangian density in terms of the renormalized quantities is
\eqs{
{\cal L} =& - \frac{1}{4} F_{\mu \nu}^{2} - \frac{1}{2 \xi^{\rm gf}} (\partial_{\mu} A^{\mu})^{2} + |\partial_{\mu} \phi|^{2} - m^{2} |\phi|^{2} + i Q {\tilde \mu}^{\epsilon/2} e A^{\mu} (\phi^{*} \partial_{\mu} \phi + Q^{2} {\tilde \mu}^{\epsilon} e^{2} A_{\mu}^{2} |\phi|^{2} - \frac{1}{4} {\tilde \mu}^{\epsilon} {\lambda} |\phi|^{4}  \\
& - \frac{1}{4} (Z_{A} - 1) F_{\mu \nu}^{2} - \frac{1}{2 \xi^{\rm gf}} (Z_{\xi^{\rm gf}} - 1) (\partial_{\mu} A^{\mu})^{2} + (Z_{\phi} - 1) |\partial_{\mu} \phi|^{2} - (Z_{m^{2}} - 1) m^{2} |\phi|^{2} \\
& + i Q (Z_{e} - 1) {\tilde \mu}^{\epsilon/2} e A^{\mu} (\phi^{*} \partial_{\mu} \phi - \partial_{\mu} \phi^{*} \phi) + Q^{2} (Z_{e^{2}} - 1) {\tilde \mu}^{\epsilon} e^{2} A_{\mu}^{2} |\phi|^{2} - \frac{1}{4} (Z_{\lambda} - 1) {\tilde \mu}^{\epsilon} \lambda |\phi|^{4}  \,.
}
The Ward-Takahashi identity warrants that $Z_{e} = Z_{\phi} = Z_{e^{2}}$, $Z_{A}$ is independent of $\xi^{\rm gf}$, $Z_{\xi^{\rm gf}} = 1$.
It follows that
\eqs{
\label{eq:scalarQED-beta}
& \beta^{\epsilon}_{e} = e \left( - \frac{1}{2} \epsilon + \frac{1}{2} \frac{d \ln Z_{A}}{d \ln \mu} \right) \,, \\
& \beta^{\epsilon}_{\lambda} = \lambda \left( - \epsilon + 2 \frac{d \ln Z_{\phi}}{d \ln \mu} - \frac{d \ln Z_{\lambda}}{d \ln \mu} \right) \,, \\
& \beta_{m^{2}} = m^{2} \left( \frac{d \ln Z_{\phi}}{d \ln \mu} - \frac{d \ln Z_{m^{2}}}{d \ln \mu} \right) \,, \\
& \beta_{\xi^{\rm gf}} = - \xi^{\rm gf} \frac{d \ln Z_{A}}{d \ln \mu} \,, \\
& \beta_{\eta} = \eta \left( \frac{d \ln Z_{\phi}}{d \ln \mu} - \frac{d \ln Z_{\eta}}{d \ln \mu} \right) \,.
}

\section{One-loop order in the scalar QED \label{sec:oneloop_app}}

One-loop functions are summarized in \cref{sec:loopfunc}.
The arguments of the one-loop functions are omit when they are obvious.
The results here are applicable to the scalar QCD by replacing $Q^{2}$ by $T(S)$ denoting the one-half of the Dynkin index of the representation for scalar fields. 

The one-loop self-energy of the gauge boson $A_{\mu}$ is given by
\eqs{
i \Pi^{\mu \nu} (p) & = (i Q e {\tilde \mu}^{\epsilon / 2})^{2} \int \frac{d^{d} \ell}{(2 \pi)^{d}} \frac{i (2 \ell + p)^{\mu} i (2 \ell + p)^{\nu}}{[\ell^{2} - m^{2}] [(\ell + p)^{2} - m^{2}]}
+ (2 i Q^{2} e^{2} g^{\mu \nu} {\tilde \mu}^{\epsilon}) \int \frac{d^{d} \ell}{(2 \pi)^{d}} \frac{i}{[\ell^{2} - m^{2}]} \\
& = i \frac{Q^{2} e^{2}}{16 \pi^{2}} \left[ \left( 4 B_{22} - 2 A \right) g^{\mu \nu} + \left( 4 B_{21}+ 4 B_{1} + B_{0}  \right) p^{\mu} p^{\nu} \right] \\
& = i (p^{2} g^{\mu \nu} - p^{\mu} p^{\nu}) \frac{Q^{2} e^{2}}{16 \pi^{2}} \frac{4}{3 p^{2}} \left[ - A + \left( m^{2} - \frac{p^{2}}{4} \right) B_{0} + m^{2} - \frac{p^{2}}{6} \right] \,.
}
The counterterm is determined to make the vacuum polarization finite,
\eqs{
\left(i \Pi^{\mu \nu} \right)^{\rm c.t.} (p) = - i (Z_{A} - 1) (p^{2} g^{\mu \nu} - p^{\mu} p^{\nu}) \,,
}
and we obtain
\eqs{
\label{eq:scalarQED-ZA}
Z_{A} - 1 = - \frac{2}{3} \frac{Q^{2} e^{2}}{16 \pi^{2}} \frac{1}{\epsilon} \,.
}
The beta function $\beta_{e}$ at the one-loop level is 
\eqs{
\label{eq:scalarQED-betae}
\beta_{e} = \frac{1}{3} \frac{Q^{2} e^{3}}{16 \pi^{2}} \,.
}
The resultant self-energy is
\eqs{
\Gamma_{2}^{\mu \nu} (p) =& - (p^{2} g^{\mu \nu} - p^{\mu} p^{\nu}) \left[ 1 + \frac{Q^{2} e^{2}}{16 \pi^{2}} \frac{4}{3 p^{2}} \left( m^{2} \int_{0}^{1} dx \ln \left( \frac{m^{2} - x (1- x) p^{2}}{m^{2}} \right) \right. \right. \\
& \left. \left. - \frac{p^{2}}{4} \int_{0}^{1} dx \ln \left( \frac{m^{2} - x (1- x) p^{2}}{\mu^{2}} \right) + \frac{p^{2}}{6} \right) \right] \,.
}

The one-loop self-energy of the scalar field $\phi$ is given by
\eqs{
i \Pi (p^{2}) & = (i Q e {\tilde \mu}^{\epsilon / 2})^{2} \int \frac{d^{d} \ell}{(2 \pi)^{d}} \frac{- i g_{\mu \nu}}{\ell^{2}} \frac{(\ell + 2 p)^{\mu} i (\ell + 2 p)^{\nu}}{(\ell + p)^{2} - m^{2}} 
+ (2 i Q^{2} e^{2} g^{\mu \nu} {\tilde \mu}^{\epsilon}) \frac{1}{2} \int \frac{d^{d} \ell}{(2 \pi)^{d}} \frac{- i g_{\mu \nu}}{\ell^{2}}
\\
& = i \frac{Q^{2} e^{2}}{16 \pi^{2}} \left[ (d - 1) A (0^{2}) - 2 p^{2} B_{1} (p^{2}; 0^{2}, m^{2}) - (3 p^{2} + m^{2}) B_{0} \right] \\
& = i \frac{Q^{2} e^{2}}{16 \pi^{2}} \left[ A (m^{2}) - 2 (p^{2} + m^{2}) B_{0} \right] \,.
}
In the first equality, we take into account the symmetric factor of $1/2$.
In the second equality, we use $g_{\mu \nu} (\ell + 2 p)^{\mu} (\ell + 2 p)^{\nu} = [(\ell + p)^{2} - m^{2}] + 2 \ell \cdot p + 3 p^{2} + m^{2}$.
The counterterm is determined to make the vacuum polarization finite,
\eqs{
i \Pi^{\rm c.t.} (p^{2}) = - i (Z_{m^{2}} - 1) m^{2} + i (Z_{\phi} - 1) p^{2} \,,
}
and we obtain
\eqs{
\label{eq:scalarQED-Z}
Z_{m^{2}} - 1 = - 2 \frac{Q^{2} e^{2}}{16 \pi^{2}} \frac{1}{\epsilon} \,, \quad Z_{\phi} - 1 = 4 \frac{Q^{2} e^{2}}{16 \pi^{2}} \frac{1}{\epsilon} \,.
}
The anomalous dimension of $\phi$ is
\eqs{
\gamma_{\phi} \equiv Z_{\phi}^{-1} \frac{d Z_{\phi}}{d \ln \mu}
= - 4 \frac{Q^{2} e^{2}}{16 \pi^{2}} \,.
}
One obtains
\eqs{
\label{eq:betam}
\beta_{m^{2}} = - 6 m^{2} \frac{Q^{2} e^{2}}{16 \pi^{2}} \,.
}
The resultant self-energy is
\eqs{
\Gamma_{2} (p^{2}) =& p^{2} - m^{2} + \frac{Q^{2} e^{2}}{16 \pi^{2}} \left[ m^{2} \left( - \int_{0}^{1} dx \ln \left( \frac{m^{2}}{\mu^{2}} \right) + 1 \right) \right. \\
& \left. + 2 (p^{2} + m^{2}) \int_{0}^{1} dx \ln \left( \frac{x m^{2} - x (1- x) p^{2}}{\mu^{2}} \right)  \right] \,.
}

\cref{fig:eta} shows one-loop diagrams contributing to $\sigma (p) \to \phi^{*} (q) \phi (k)$ in the scalar QED.
\cref{fig:eta} (a) gives
\eqs{
\label{eq:etaa}
i {\cal M}_{(a)}^{\rm loop} & = 2 i (m^{2} - \eta p^{2}) (i Q e {\tilde \mu}^{\epsilon/2})^{2} \int \frac{d^{d} \ell}{(2 \pi)^{d}} \frac{i (\ell - k)^{\nu}}{\ell^{2} - m^{2}} \frac{- i g_{\mu\nu}}{(\ell + k)^{2}} \frac{i (\ell + q + p)^{\mu}}{(\ell + p)^{2} - m^{2}} \\
& = 2 i (m^{2} - \eta p^{2}) \frac{Q^{2} e^{2}}{16 \pi^{2}} \left[ B_{0} (p^{2}; m^{2}, m^{2}) + 2 (q - k)^{\mu} C_{\mu} (k^{2}, q^{2}, p^{2}; m^{2}, 0, m^{2}) - 2 k \cdot (q + k) C_{0} \right] \\
& = 2 i (m^{2} - \eta p^{2}) \frac{Q^{2} e^{2}}{16 \pi^{2}} \left[ B_{0} (k^{2}; m^{2}, 0^{2}) + B_{0} (q^{2}; 0^{2}, m^{2}) - B_{0} (p^{2}; m^{2}, m^{2}) - (2 p^{2} - k^{2} - q^{2} - 2 m^{2}) C_{0} \right] \,.
}
In the second equality, we use $(\ell - k) \cdot (\ell + p + q) = (\ell + k)^{2} + 2 \ell \cdot (q - k) - 2 k \cdot (q + k)$.

\cref{fig:eta} (b) gives
\eqs{
i {\cal M}_{(b)}^{\rm loop} &= i \epsilon (2 i Q^{2} e^{2} {\tilde \mu}^{\epsilon}) \frac{1}{2} \int \frac{d^{d} \ell}{(2 \pi)^{d}} \left[- \ell \cdot (\ell + p) g^{\mu \nu} + \ell^{\mu} (\ell + p)^{\nu} \right] \frac{- i g_{\mu \rho}}{\ell^{2}} \frac{- i g_{\nu \sigma}}{(\ell + p)^{2}} g^{\rho \sigma} \\
&= - (d - 1) i \epsilon \frac{Q^{2} e^{2}}{16 \pi^{2}} \left[A (0^{2}) + p^{\mu} B_{\mu} (p^{2}; 0^{2}, 0^{2}) \right] \\
& = \frac{3 - \epsilon}{2} i \epsilon \frac{Q^{2} e^{2}}{16 \pi^{2}} p^{2} B_{0} (p^{2}; 0^{2}, 0^{2}) \,.
}
In the first equality, we take into account the symmetric factor of $1/2$.
In the second equality, we use $g_{\mu \rho} g_{\nu \sigma} g^{\rho \sigma} \left[- \ell \cdot (\ell + p) g^{\mu \nu} + \ell^{\mu} (\ell + p)^{\nu} \right] = \ell^{2} + \ell \cdot p$.

\cref{fig:eta} (c) gives
\eqs{
i {\cal M}_{(c)}^{\rm loop} =& i \epsilon (i Q e {\tilde \mu}^{\epsilon/2})^{2} \int \frac{d^{d} \ell}{(2 \pi)^{d}} \left[- \ell \cdot (\ell + p) g^{\mu \nu} + \ell^{\mu} (\ell + p)^{\nu} \right] 
\frac{- i g_{\mu \rho}}{\ell^{2}} \frac{(- \ell - 2 k)^{\rho} i (- \ell - k + q)^{\sigma}}{(\ell + k)^{2} - m^{2}} \frac{- i g_{\nu \sigma}}{(\ell + p)^{2}} \\
=& - 4 i \epsilon \frac{Q^{2} e^{2}}{16 \pi^{2}} \left[ (k \cdot q) B_{0} (q^{2}; m^{2}, 0^{2}) - k^{\mu} q^{\nu} C_{\mu \nu} (k^{2}, q^{2}, p^{2}; 0^{2}, m^{2}, 0^{2}) + ( (k \cdot q) k^{\mu} - k^{2} q^{\mu} ) C_{\mu} \right] \\
=& - i \epsilon \frac{Q^{2} e^{2}}{16 \pi^{2}} \left[ A (m^{2}) + (k^{2} + 3 k \cdot q + q^{2} + m^{2}) B_{0} (p^{2}; 0^{2}, 0^{2}) - (q^{2} + m^{2}) B_{0} (q^{2}; m^{2}, 0^{2}) \right. \\
& \left.  - (k^{2} + m^{2}) B_{0} (k^{2}; 0^{2}, m^{2}) + (4 m^{2} k \cdot q + (q^{2} + m^{2}) (k^{2} + m^{2}) ) C_{0} \right] \,.
}
In the second equality, we use 
\eqs{
& g_{\mu \rho} g_{\nu \sigma} \left[- \ell \cdot (\ell + p) g^{\mu \nu} + \ell^{\mu} (\ell + p)^{\nu} \right] (\ell + 2 k)^{\rho} (\ell + k - q)^{\sigma} \\
& = 4 \left[ (k \cdot q) \ell^{2} - (\ell \cdot k) (\ell \cdot q) + (k \cdot q) (\ell \cdot k) - k^{2}   (\ell \cdot q) \right] \,.
}
Here $\ell^{4}$ and $\ell^{3}$ terms vanish because
\eqs{
\left[- \ell \cdot (\ell + p) g^{\mu \nu} + \ell^{\mu} (\ell + p)^{\nu} \right] \ell_{\nu} =  (\ell + p)_{\mu} \left[- \ell \cdot (\ell + p) g^{\mu \nu} + \ell^{\mu} (\ell + p)^{\nu} \right] = 0 \,.
}

We consider the renormalization of a composite operator $|\phi|^{2}$ (mass operator).
The renormalized mass operator is written in terms of bare/renormalized fields as follows:
\eqs{
|\phi_{0}|^{2} = Z_{\phi} |\phi|^{2} = Z_{\phi^{2}} [|\phi|^{2}] \,.
}
Here, a square bracket denotes renormalized composite operators.
\cref{fig:eta} (a), with the insertion of $|\phi|^{2}$ instead of $T^{\mu}_{~ \mu} \supset 2 m^{2} |\phi|^{2} + 2 \eta \partial^{2} |\phi|^{2}$, gives the one-loop diagram:
\eqs{
i {\cal M}^{\rm loop} =& i (i Q e {\tilde \mu}^{\epsilon/2})^{2} \int \frac{d^{d} \ell}{(2 \pi)^{d}} \frac{i (\ell - k)^{\nu}}{\ell^{2} - m^{2}} \frac{- i g_{\mu\nu}}{(\ell + k)^{2}} \frac{i (\ell + q + p)^{\mu}}{(\ell + p)^{2} - m^{2}} \\
=& i \frac{Q^{2} e^{2}}{16 \pi^{2}} \left[ B_{0} (k^{2}; m^{2}, 0^{2}) + B_{0} (q^{2}; 0^{2}, m^{2}) - B_{0} (p^{2}; m^{2}, m^{2}) \right. \\
& \left. - (2 p^{2} - k^{2} - q^{2} - 2 m^{2}) C_{0} (k^{2}, q^{2}, p^{2}; 0^{2}, m^{2}, 0^{2}) \right] \,.
}
The divergent part is
\eqs{
\left( i {\cal M} \right)^{\rm pole}_{\rm of \, \epsilon} & =
2 i \frac{Q^{2} e^{2}}{16\pi^{2}} \frac{1}{\epsilon} \,.
}
This divergence is canceled by the counterterm,
\eqs{
i {\cal M}^{\rm c.t.} = i (Z_{\phi^{2}}^{-1} Z_{\phi} - 1) \,,
}
and thus
\eqs{
Z_{\phi^{2}}^{-1} Z_{\phi} - 1 = - 2 \frac{Q^{2} e^{2}}{16\pi^{2}} \frac{1}{\epsilon} \,.
}
Since $Z_{\phi}$ at the one-loop level is given in \cref{eq:scalarQED-Z}, one obtains $Z_{\phi^{2}}$ and its anomalous dimension as follows:
\eqs{
\label{eq:scalarQED-Zsqed}
Z_{\phi^{2}} - 1 = 6 \frac{Q^{2} e^{2}}{16\pi^{2}} \frac{1}{\epsilon} \,, \quad \gamma_{\phi^{2}} = \frac{d \ln Z_{\phi^{2}}}{d \ln \mu} = - 6 \frac{Q^{2}  e^{2}}{16\pi^{2}} \,.
}

\subsection{Summary of one-loop functions \label{sec:loopfunc}}

One-loop functions are based on Refs.~\cite{tHooft:1978jhc, Passarino:1978jh} (see also Appendix F of Ref.~\cite{Logan:1999if}).

The one-point integral is defined as
\eqs{
{\tilde \mu}^{\epsilon} \int \frac{d^{d} \ell}{(2 \pi)^{d}} \frac{1}{\ell^{2} - m^{2}} = \frac{i}{16 \pi^{2}} A (m^{2}) \,.
}
The explicit form is
\eqs{
A (m^{2}) = m^{2} \left( \frac{2}{\epsilon} - \ln \left( \frac{m^{2}}{\mu^{2}} \right) + 1 \right) \,.
}

Two-point integrals are defined as
\eqs{
{\tilde \mu}^{\epsilon} \int \frac{d^{d} \ell}{(2 \pi)^{d}} \frac{1; \ell_{\mu}; \ell_{\mu} \ell_{\nu}}{[\ell^{2} - m_{1}^{2}] [(\ell+p)^{2} - m_{2}^{2}]} = \frac{i}{16 \pi^{2}} B_{0; \mu; \mu \nu} (p^{2}; m_{1}^{2}, m_{2}^{2}) \,,
}
where
\eqs{
& B_{\mu} = p_{\mu} B_{1} \,, \\
& B_{\mu \nu} = g_{\mu \nu} B_{22} + p_{\mu} p_{\nu} B_{21} \,.
}
They can be reduced to $A$ and $B_{0}$ as
\eqs{
& B_{1} (p^{2}; m_{1}^{2}, m_{2}^{2}) = \frac{1}{2 p^{2}} \left[ A (m_{1}^{2}) - A (m_{2}^{2}) - (p^{2} + m_{1}^{2} - m_{2}^{2}) B_{0} \right] \,, \\
& B_{22} (p^{2}; m_{1}^{2}, m_{2}^{2}) = \frac{1}{6} \left[ A (m_{2}^{2}) + 2 m_{1}^{2} B_{0} + (p^{2} + m_{1}^{2} - m_{2}^{2}) B_{1} + m_{1}^{2} + m_{1}^{2} - \frac{p^{2}}{3} \right] \,, \\
& B_{21} (p^{2}; m_{1}^{2}, m_{2}^{2}) = \frac{1}{3 p^{2}} \left[ A (m_{2}^{2}) - m_{1}^{2} B_{0} - 2 (p^{2} + m_{1}^{2} - m_{2}^{2}) B_{1} - \frac{m_{1}^{2} + m_{1}^{2}}{2} + \frac{p^{2}}{6} \right] \,.
}
The explicit form with a Feynman parameter integral is
\eqs{
B_{0} = \frac{2}{\epsilon} - \int^{1}_{0} dx \ln \left( \frac{x^{2} p^{2} - x (p^{2} + m_{1}^{2} - m_{2}^{2}) + m_{1}^{2} - i \epsilon_{\rm ad}}{\mu^{2}} \right) \,.
}


Three-point integrals are defined as
\eqs{
{\tilde \mu}^{\epsilon} \int \frac{d^{d} \ell}{(2 \pi)^{d}} \frac{1; \ell_{\mu}; \ell_{\mu} \ell_{\nu}}{[\ell^{2} - m_{1}^{2}] [(\ell + k)^{2} - m_{2}^{2}] [(\ell + k + q)^{2} - m_{3}^{2}]} = \frac{i}{16 \pi^{2}} C_{0; \mu; \mu \nu} (k^{2}, q^{2}, p^{2}; m_{1}^{2}, m_{2}^{2}, m_{3}^{2}) \,,
}
where $p + q + k = 0$ and
\eqs{
& C_{\mu} = k_{\mu} C_{11} + q_{\mu} C_{12} \,, \\
& C_{\mu \nu} = g_{\mu \nu} C_{24} + k_{\mu} k_{\nu} C_{21} + q_{\mu} q_{\nu} C_{22} + \left( k_{\mu} q_{\nu} +  q_{\mu} k_{\nu} \right) C_{23} \,.
}
They can be reduced to $A$, $B_{0}$, and $C_{0}$ as
\eqs{
& C_{11} (k^{2}, q^{2}, p^{2}; m_{1}^{2}, m_{2}^{2}, m_{3}^{2}) = \frac{1}{k^{2} q^{2} - (k \cdot q)^{2}} \left[ q^{2} R_{1} - (k \cdot q) R_{2} \right] \,, \\
& C_{12} (k^{2}, q^{2}, p^{2}; m_{1}^{2}, m_{2}^{2}, m_{3}^{2}) = \frac{1}{k^{2} q^{2} - (k \cdot q)^{2}} \left[ - (k \cdot q) R_{1} + k^{2} R_{2} \right] \,, \\
& C_{24} (k^{2}, q^{2}, p^{2}; m_{1}^{2}, m_{2}^{2}, m_{3}^{2}) \\
& = \frac{1}{4} \left[ B_{0} (q^{2}; m_{2}^{2}, m_{3}^{2}) + (k^{2} + m_{1}^{2} - m_{2}^{2}) C_{11} + (p^{2} - k^{2} + m_{2}^{2} - m_{3}^{2}) C_{12} + 2 m_{1}^{2} C_{0} + 1 \right] \,, \\
& C_{21} (k^{2}, q^{2}, p^{2}; m_{1}^{2}, m_{2}^{2}, m_{3}^{2}) = \frac{1}{k^{2} q^{2} - (k \cdot q)^{2}} \left[ q^{2} R_{3} - (k \cdot q) R_{5} \right] \,, \\
& C_{23} (k^{2}, q^{2}, p^{2}; m_{1}^{2}, m_{2}^{2}, m_{3}^{2}) = \frac{1}{k^{2} q^{2} - (k \cdot q)^{2}} \left[ - (k \cdot q) R_{3} + k^{2} R_{5} \right] = \frac{1}{k^{2} q^{2} - (k \cdot q)^{2}} \left[ q^{2} R_{4} - (k \cdot q) R_{6} \right] \,, \\
& C_{22} (k^{2}, q^{2}, p^{2}; m_{1}^{2}, m_{2}^{2}, m_{3}^{2}) = \frac{1}{k^{2} q^{2} - (k \cdot q)^{2}} \left[ - (k \cdot q) R_{4} + k^{2} R_{6} \right] \,,
}
where
\eqs{
& R_{1} (k^{2}, q^{2}, p^{2}; m_{1}^{2}, m_{2}^{2}, m_{3}^{2}) = \frac{1}{2} \left[ B_{0} (p^{2}; m_{1}^{2}, m_{3}^{2}) - B_{0} (q^{2}; m_{2}^{2}, m_{3}^{2}) - (k^{2} + m_{1}^{2} - m_{2}^{2}) C_{0} \right] \,, \\ 
& R_{2} (k^{2}, q^{2}, p^{2}; m_{1}^{2}, m_{2}^{2}, m_{3}^{2}) = \frac{1}{2} \left[ B_{0} (k^{2}; m_{1}^{2}, m_{2}^{2}) - B_{0} (p^{2}; m_{1}^{2}, m_{3}^{2}) + (p^{2} - k^{2} + m_{2}^{2} - m_{3}^{2}) C_{0} \right] \,, \\ 
& R_{3} (k^{2}, q^{2}, p^{2}; m_{1}^{2}, m_{2}^{2}, m_{3}^{2}) = - C_{24} - \frac{1}{2} \left[ (k^{2} + m_{1}^{2} - m_{2}^{2}) C_{11} - B_{1} (p^{2}; m_{1}^{2}, m_{3}^{2}) - B_{0} (q^{2}; m_{2}^{2}, m_{3}^{2}) \right] \,, \\ 
& R_{4} (k^{2}, q^{2}, p^{2}; m_{1}^{2}, m_{2}^{2}, m_{3}^{2}) = - \frac{1}{2} \left[ (k^{2} + m_{1}^{2} - m_{2}^{2}) C_{12} - B_{1} (p^{2}; m_{1}^{2}, m_{3}^{2}) + B_{1} (q^{2}; m_{2}^{2}, m_{3}^{2}) \right] \,, \\ 
& R_{5} (k^{2}, q^{2}, p^{2}; m_{1}^{2}, m_{2}^{2}, m_{3}^{2}) = - \frac{1}{2} \left[ (p^{2} - k^{2} + m_{2}^{2} - m_{3}^{2}) C_{11} - B_{1} (k^{2}; m_{1}^{2}, m_{2}^{2}) + B_{1} (p^{2}; m_{1}^{2}, m_{3}^{2}) \right] \,, \\ 
& R_{6} (k^{2}, q^{2}, p^{2}; m_{1}^{2}, m_{2}^{2}, m_{3}^{2}) = - C_{24} - \frac{1}{2} \left[ (p^{2} - k^{2} + m_{2}^{2} - m_{3}^{2}) C_{12} + B_{1} (p^{2}; m_{1}^{2}, m_{3}^{2}) \right] \,.
}
The explicit form with Feynman parameter integrals is%
\footnote{
The version with $x \to 1 - x$, i.e., $\int_{0}^{1} dx \int_{0}^{1 - x} dy \dots$ may also be familiar to readers.
}
\eqs{
C_{0} =& - \int_{0}^{1} dx \int_{0}^{x} dy \\ 
& \times \frac{1}{x^{2} p^{2} + y^{2} q^{2} - x y (p^{2} - k^{2} + q^{2}) - x (p^{2} + m_{1}^{2} - m_{3}^{2}) + y (p^{2} - k^{2} + m_{2}^{2} - m_{3}^{2}) + m_{1}^{2} - i \epsilon_{\rm ad}} \,.
}


\section{Two-loop order in the scalar QED \label{sec:twoloop_app}}

We take a closer look at the two-loop-order contributions shown in \cref{fig:twoloop} with the insertion of $T^{\mu}_{~ \mu} \supset - (1/4) \epsilon F^{2}_{\mu \nu}$.
There are two types of diagrams: two-loop diagrams without the insertion of counterterms and one-loop diagrams with the insertion of one-loop counterterms (filled dots in \cref{fig:twoloop}).
We consider subdivergences of the former two-loop diagrams.
There are two subdivergences in each two-loop diagram, depending on which loop momentum gets large.
In the following, we list the corresponding counterterm contributions and check that their sum reproduces that of one-loop diagrams with one-loop counterterms in \cref{fig:twoloop}.
We take the massless limit of the scalar $\phi$, $m \to 0$, since we focus on contributions to the non-minimal coupling.

\begin{figure}
	\centering
	\includegraphics[width=0.75\linewidth]{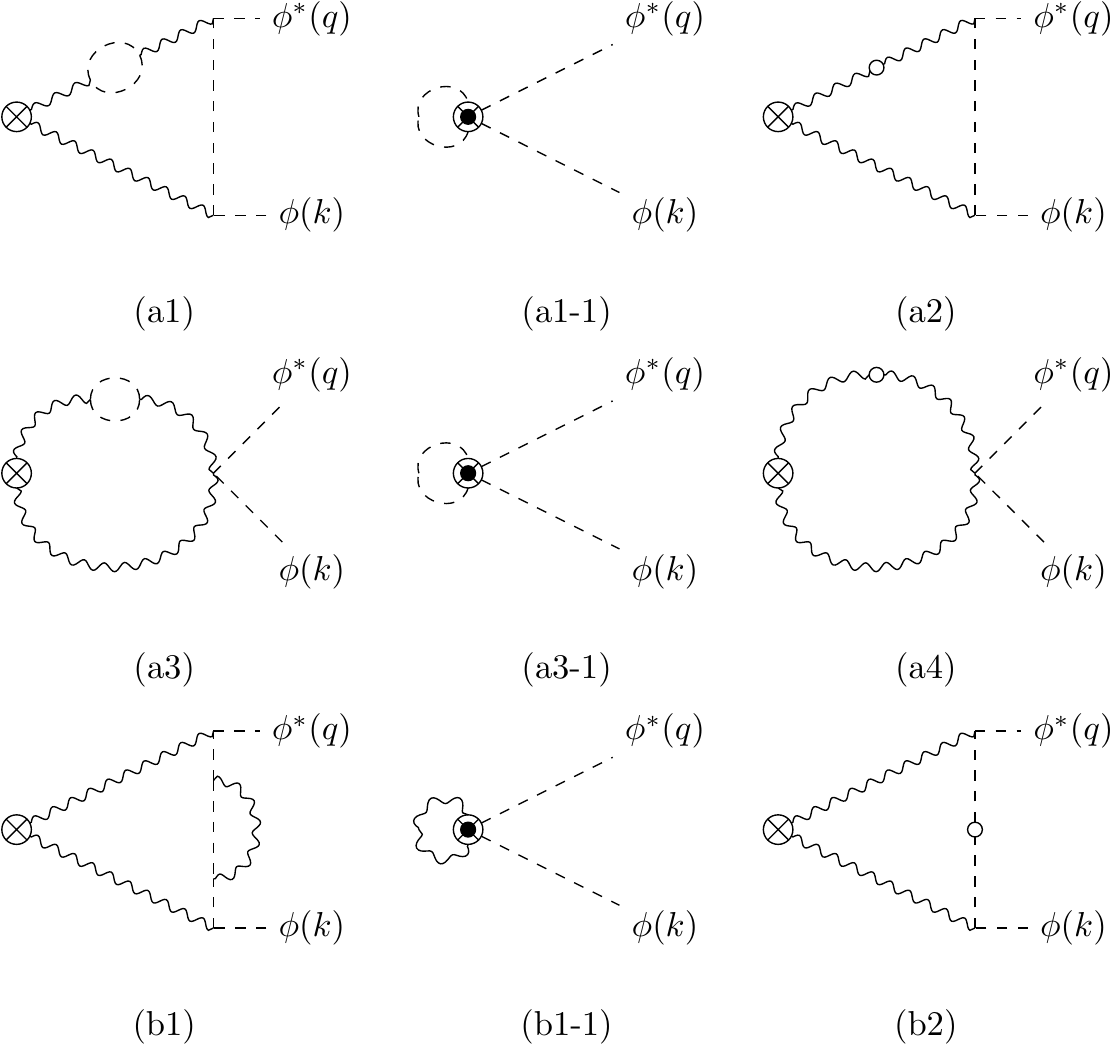}
	\caption{Two-loop diagrams (left column) for scalaron decay $\sigma (p) \to \phi^{*} (q) \phi (k)$ among the type-(a) and (b) diagrams in \cref{fig:twoloop}.
	For the first two rows, there are contributions also from diagrams with internal photons being exchanged.
	Crossed dots denote insertion of the energy-momentum tensor, $T^{\mu}_{~ \mu} \supset - (1/4) \epsilon F^{2}_{\mu \nu}$.
	The corresponding counterterm diagrams are shown in the middle and right columns.
	The middle-column diagrams are with the insertion of the counterterms in $T^{\mu}_{~ \mu}$ (filled-crossed dots).
	The right-column diagrams are with the insertion of the counterterms for propagators (unfilled dots).
	In each row, non-local divergences are canceled.}
	\label{fig:BPHZ_AB}
\end{figure}

\begin{figure}
	\centering
	\includegraphics[width=0.75\linewidth]{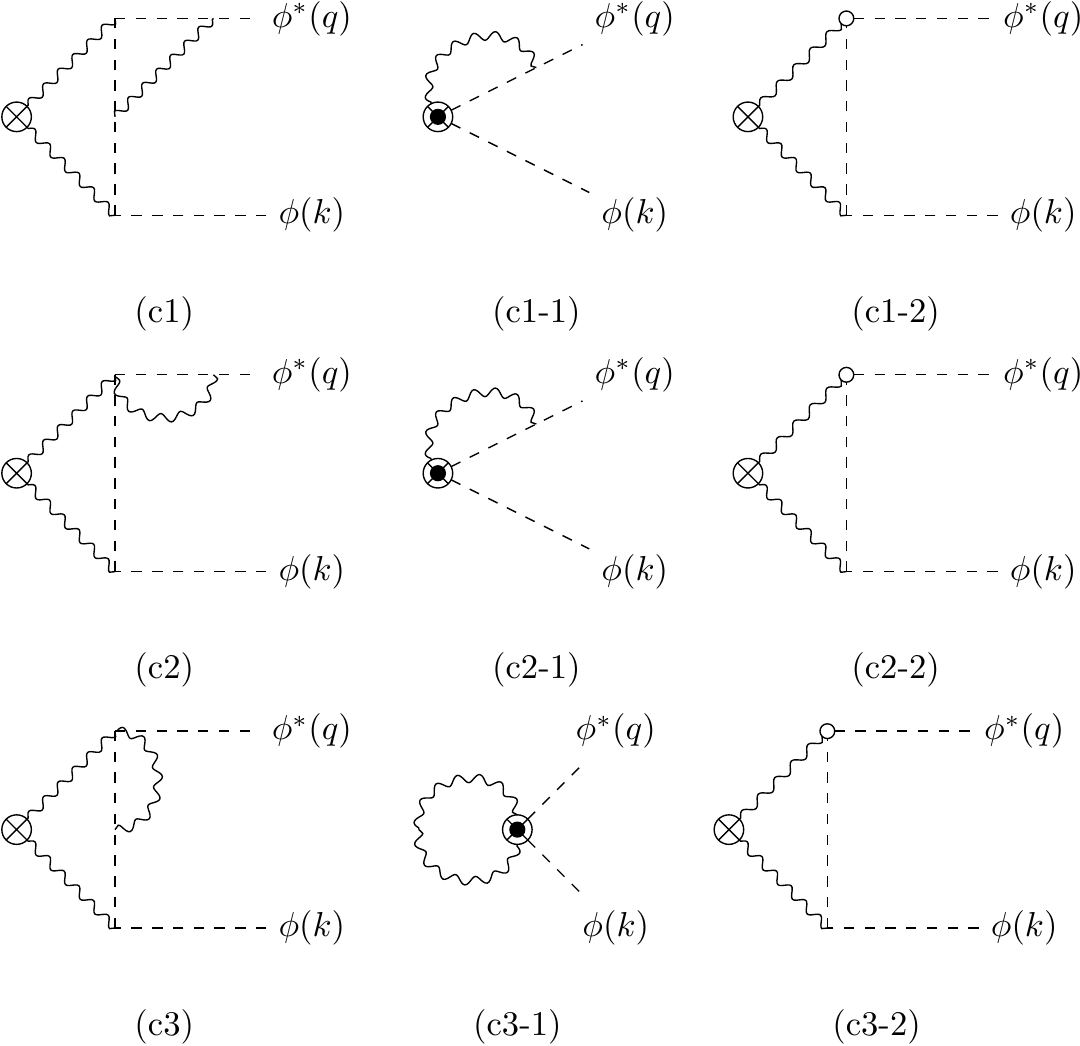}
	\caption{The same as \cref{fig:BPHZ_AB} but for the type-(c) diagrams.
	There are contributions also from diagrams with internal photons being exchanged.}
	\label{fig:BPHZ_C}
\end{figure}

\begin{figure}
	\centering
	\includegraphics[width=0.75\linewidth]{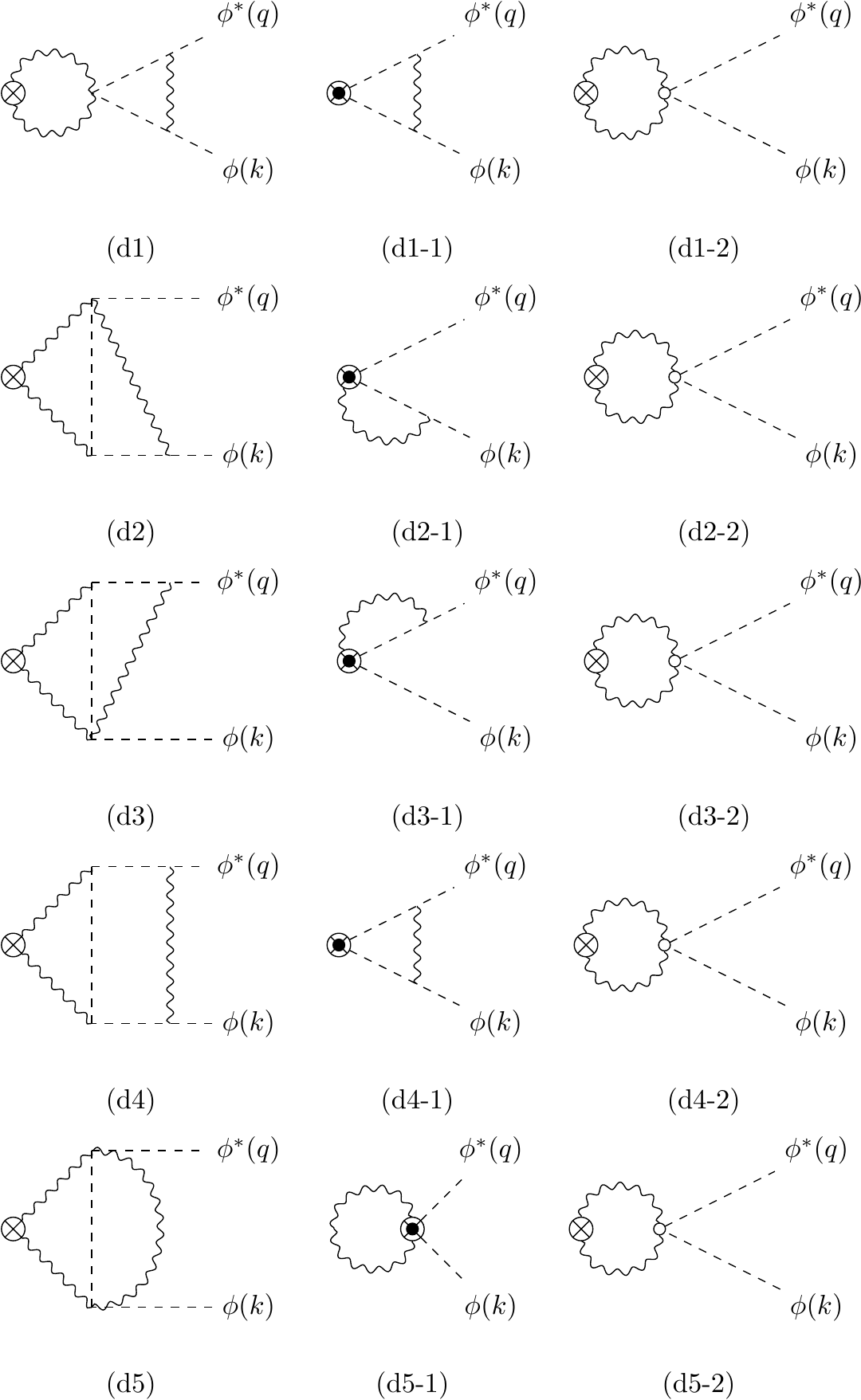}
	\caption{The same as \cref{fig:BPHZ_AB} but for the type-(d) diagrams.}
	\label{fig:BPHZ_D}
\end{figure}

\cref{fig:BPHZ_AB,fig:BPHZ_C,fig:BPHZ_D} show the two-loop diagrams, type-(a), (b), (c), and (d) diagrams in \cref{fig:twoloop}, and their subdivergences.
We put the two-loop diagrams on the left columns in these figures, while the corresponding counterterm diagrams on the middle and right columns.
The middle-column diagrams in \cref{fig:BPHZ_AB,fig:BPHZ_C,fig:BPHZ_D} are with the insertion of counterterms in $T^{\mu}_{~ \mu}$ (filled-crossed dots).%
\footnote{
Precisely speaking, they are not counterterms since they are finite due to $\epsilon$ in $T^{\mu}_{~ \mu} \supset - (1/4) \epsilon F^{2}_{\mu \nu}$.
}
On the other hand, the right-column diagrams are with the insertion of counterterms for propagators, $\phi^\ast \phi A$ vertex, and $\phi^\ast \phi A A$ vertex (unfilled dots).
We give the detailed computation of the subdivergences in \cref{sec:oneloop_calc}.

Among the type-(a) and (b) diagrams in \cref{fig:twoloop}, two-loop diagrams are (a1), (a3) and (b1) which are shown in the left column of \cref{fig:BPHZ_AB}.
For rows of (a1) and (a3), there are contributions also from diagrams with internal photons being exchanged.
In each row, non-local divergences are canceled in light of the BPHZ theorem~\cite{Bogoliubov:1957gp, Hepp:1966eg, Zimmermann:1969jj}.
The filled-crossed dots in the middle-column diagrams correspond to the insertion of counterterms in $T^{\mu}_{~ \mu}$: %
(a1-1, a3-1) $\leftrightarrow |\phi|^{4}$ and (b1-1) $\leftrightarrow |\phi^{2}| A_{\mu}^{2}$.
They vanish in the dimensional regularization, since their loop integral is quadratically divergent and contains only massless particles.
The right-column diagrams reproduce one-loop diagrams with one-loop counterterms in \cref{fig:twoloop}: (a2, a4) $\leftrightarrow Z_{A} - 1$ [see \cref{eq:scalarQED-ZA}] and (b2) $\leftrightarrow Z_{\phi} - 1$ [see \cref{eq:scalarQED-Z}], respectively.
In other words, there is no difference between the unfilled dots in \cref{fig:BPHZ_AB} and filled ones in \cref{fig:twoloop}.

\begin{figure}
	\centering
	\includegraphics[width=0.8\linewidth]{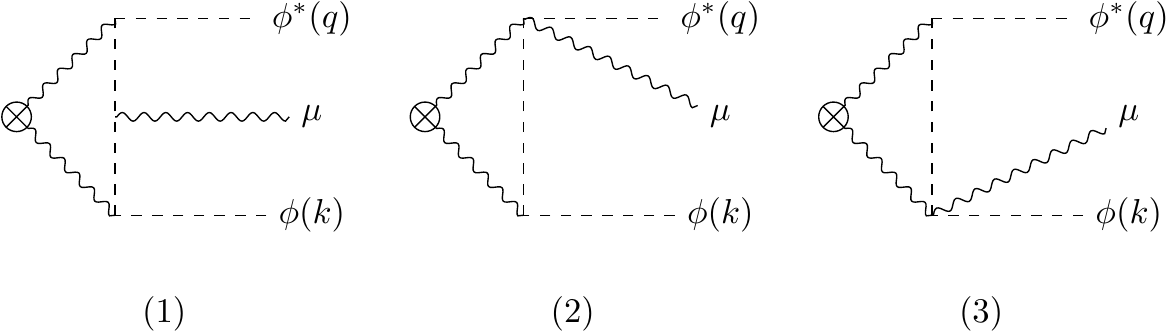}
	\caption{Type-2) one-loop diagrams with outgoing $\phi^{*} (q) \phi (k) A_{\mu}$ (not necessarily on-shell). 	
	Crossed dots denote insertion of $- (1/4) \epsilon F^{2}_{\mu \nu}$ with the incoming momentum of $p$.}
	\label{fig:BPHZ_eom}
\end{figure}

\begin{figure}
	\centering
	\includegraphics[width=0.7\linewidth]{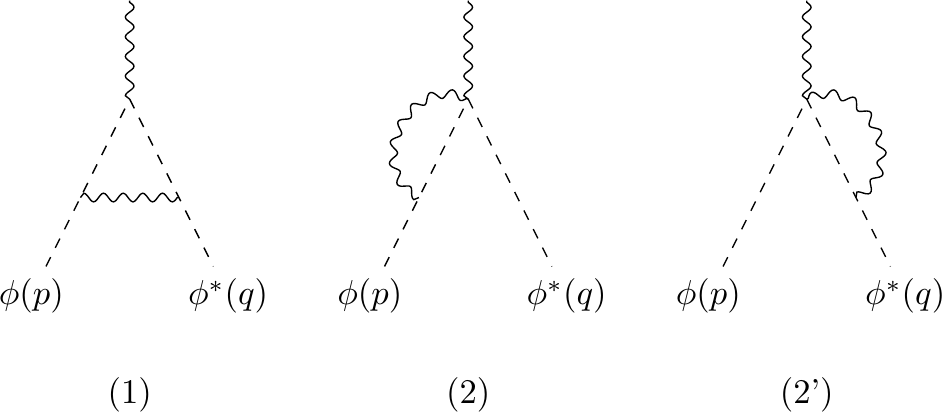}
	\caption{One-loop corrections to the $\phi^{*} (q) \phi (p) A_{\mu}$ vertex.
	Note that $p$ is incoming, while $q$ is out-going.}
	\label{fig:oneloop_v3}
\end{figure}

Among the type-(c) diagrams in \cref{fig:twoloop}, two-loop diagrams are (c1)-(c3) which are shown in the left column of \cref{fig:BPHZ_C}.
For each row, there are contributions also from diagrams with internal photons being exchanged.
The counterterm [except for (c3-1)] (denoted by filled-crossed dots in the middle-column diagrams) cancels with the divergence of the corresponding diagram in \cref{fig:BPHZ_eom}: (c1-1) $\leftrightarrow$ \cref{fig:BPHZ_eom} (1), (c2-1) $\leftrightarrow$ \cref{fig:BPHZ_eom} (2) [and (3)], and (c3-1) $\leftrightarrow |\phi^{2}| A_{\mu}^{2}$.
The (c1-1) diagram vanishes because \cref{fig:BPHZ_eom} (1) is finite and does not have any counterterm.
The (c2-1) diagram is divergent and may cause a possible issue because there is no corresponding counterterm diagram in \cref{fig:twoloop}.
We see that this cancels with other diagrams shortly below.
The (c3-1) diagram vanishes, because of the quadratically divergent loop integral.
The counterterm (unfilled dots in the right-column diagrams) cancels the divergence of each diagram in \cref{fig:oneloop_v3}: (c1-2) $\leftrightarrow$ \cref{fig:oneloop_v3} (1), (c2-2) $\leftrightarrow$ \cref{fig:oneloop_v3} (2'), and (c3-2) $\leftrightarrow$ \cref{fig:oneloop_v3} (2).
In \cref{sec:oneloop_calc}, we determine the counterterms by evaluating the one-loop corrections to the $\phi^{*} (q) \phi (p) A_{\mu}$ vertex as shown in \cref{fig:oneloop_v3}.
Here $p$ is an incoming momentum, while $q$ is out-going.
The results are
\eqs{
& i \Gamma_{3;1;{\rm c.t.}}^{\mu} (p, q) = (i Q e {\tilde \mu}^{\epsilon / 2}) (-2) (p + q)^{\mu} \frac{Q^{2} e^{2}}{16 \pi^{2}} \frac{1}{\epsilon} \,, \\
& i \Gamma_{3;2;{\rm c.t.}}^{\mu} (p, q) = (i Q e {\tilde \mu}^{\epsilon / 2}) 6 p^{\mu} \frac{Q^{2} e^{2}}{16 \pi^{2}} \frac{1}{\epsilon} \,, \quad 
i \Gamma_{3;2';{\rm c.t.}}^{\mu} (p, q) = (i Q e {\tilde \mu}^{\epsilon / 2}) 6 q^{\mu} \frac{Q^{2} e^{2}}{16 \pi^{2}} \frac{1}{\epsilon} \,.
}
The sum of the counterterms is nothing but the one-loop counterterm of of the $\phi^{*} \phi A$ vertex: $(i Q e {\tilde \mu}^{\epsilon / 2}) (Z_{e} - 1) (p + q)^{\mu}$, where we determine
\eqs{
Z_{e} - 1 = 4 \frac{Q^{2} e^{2}}{16 \pi^{2}} \frac{1}{\epsilon} \,,
}
from the Ward-Takahashi identity, $Z_{e} = Z_{\phi}$ [see \cref{eq:scalarQED-Z}].
Thus, the right-column diagrams in \cref{fig:BPHZ_C} reproduce the one-loop diagram (c4) in \cref{fig:twoloop}.

\begin{figure}
	\centering
	\includegraphics[width=1.0\linewidth]{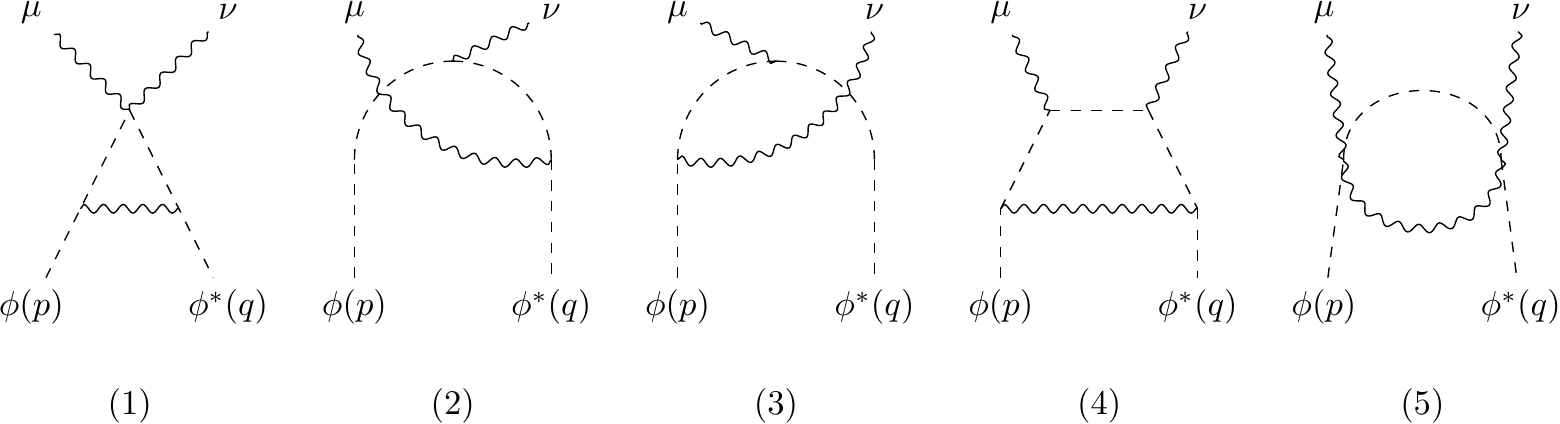}
	\caption{One-loop corrections to the $\phi^{*} (q) \phi (p) A_{\mu} (k - p) A_{\nu}$ vertex.
	Note that $p$ and $k - p$ are incoming, while $q$ is out-going.
	For (2)-(5), there are contributions with the external photons being exchanged.}
	\label{fig:oneloop_v4}
\end{figure}

Among the type-(d) diagrams in \cref{fig:twoloop}, two-loop diagrams are (d1)-(d5) which are shown in the left column of \cref{fig:BPHZ_D}.
The counterterm in the middle-column diagrams [except for (d5-1)] (denoted by filled-crossed dots) cancels with the divergence of the corresponding diagram in \cref{fig:eta} and \cref{fig:BPHZ_eom}: (d1-1) $\leftrightarrow$ \cref{fig:eta} (b), (d2-1) $\leftrightarrow$ \cref{fig:BPHZ_eom} (2), (d3-1) $\leftrightarrow$ \cref{fig:BPHZ_eom} (3), (d4-1) $\leftrightarrow$ \cref{fig:eta} (c), and (d5-1) $\leftrightarrow |\phi^{2}| A_{\mu}^{2}$.
The (d1-1)-(d4-1) diagrams are divergent, and may cause a possible issue because there is no corresponding counterterm diagram in \cref{fig:twoloop}.
We see that this cancels with other diagrams shortly below.
The (d5-1) diagram vanishes, because of the quadratically divergent loop integral.
The counterterm in the right-column diagrams (unfilled dots) cancels the divergence of each diagram in \cref{fig:oneloop_v4}: (d1-2) $\leftrightarrow$ \cref{fig:oneloop_v4} (1), (d2-2) $\leftrightarrow$ (3), (d3-2) $\leftrightarrow$ \cref{fig:oneloop_v4} (2), (d4-2) $\leftrightarrow$ \cref{fig:oneloop_v4} (4), and (d5-2) $\leftrightarrow$ \cref{fig:oneloop_v4} (5).
In \cref{sec:oneloop_calc}, we evaluate the $\phi^{*} (q) \phi (p) A_{\mu} (k - p) A_{\nu}$ vertex as shown in \cref{fig:oneloop_v4}.
Here $p$ and $k - p$ are incoming momenta, while $q$ is out-going.
For \cref{fig:oneloop_v4} (2)-(5), there are contributions with the external photons being exchanged and thus doubled.
The resultant counterterms are
\eqs{
& i \Gamma_{4;1;{\rm c.t.}}^{\mu \nu} = (2 i Q^{2} e^{2} {\tilde \mu}^{\epsilon}) (-2) g^{\mu \nu} \frac{Q^{2} e^{2}}{16 \pi^{2}} \frac{1}{\epsilon} \,, \quad 
i \Gamma_{4;2;{\rm c.t.}}^{\mu \nu} = i \Gamma_{4;3;{\rm c.t.}}^{\mu \nu} = (2 i Q^{2} e^{2} {\tilde \mu}^{\epsilon}) (-2) g^{\mu \nu} \frac{Q^{2} e^{2}}{16 \pi^{2}} \frac{1}{\epsilon} \,, \\
& i \Gamma_{4;4;{\rm c.t.}}^{\mu \nu} = (2 i Q^{2} e^{2} {\tilde \mu}^{\epsilon}) 2 g^{\mu \nu} \frac{Q^{2} e^{2}}{16 \pi^{2}} \frac{1}{\epsilon} \,, \quad 
i \Gamma_{4;5;{\rm c.t.}}^{\mu \nu} = (2 i Q^{2} e^{2} {\tilde \mu}^{\epsilon}) 8 g^{\mu \nu} \frac{Q^{2} e^{2}}{16 \pi^{2}} \frac{1}{\epsilon} \,.
}
The sum of the counterterms is nothing but the one-loop counterterm of of the $\phi^{*} \phi A A$ vertex: $(2 i Q^2 e^2 {\tilde \mu}^{\epsilon}) (Z_{e^{2}} - 1) g^{\mu \nu}$, where we determine
\eqs{
Z_{e^{2}} - 1 = 4 \frac{Q^{2} e^{2}}{16 \pi^{2}} \frac{1}{\epsilon} \,,
}
from the Ward-Takahashi identity, $Z_{e^{2}} = Z_{\phi}$ [see \cref{eq:scalarQED-Z}].
Thus, the right-column diagrams in \cref{fig:BPHZ_D} reproduce the one-loop diagram (d6) in \cref{fig:twoloop}.

In summary, the right-column diagrams in \cref{fig:BPHZ_AB,fig:BPHZ_C,fig:BPHZ_D} reproduce the one-loop diagrams with the one-loop counterterms in \cref{fig:twoloop}.
There are no more diagrams in \cref{fig:twoloop}, and thus the middle-column diagrams in \cref{fig:BPHZ_AB,fig:BPHZ_C,fig:BPHZ_D} should also be zero.
On the other hand, the (c2-1) (also the one with internal photons being exchanged) and (d1-1)-(d4-1) diagrams are each divergent.

The key is that the counterterms in $T^{\mu}_{~ \mu}$ form the equation of motion.
Each counterterm cancels the divergence of each diagram in \cref{fig:eta} (b) and (c) and \cref{fig:BPHZ_eom}: (c2-1) $\leftrightarrow$ \cref{fig:BPHZ_eom} (2) [and (3)], (d1-1) $\leftrightarrow$ \cref{fig:eta} (b), (d2-1) $\leftrightarrow$ \cref{fig:BPHZ_eom} (2), (d3-1) $\leftrightarrow$ \cref{fig:BPHZ_eom} (3), and (d4-1) $\leftrightarrow$ \cref{fig:eta} (c).
First, \cref{fig:eta} (b) and (c) [see \cref{eq:M_loop_b_c}] and their counterterms are
\eqs{
i {\cal M}_{2;(b); {\rm c.t.}} = - 3 i \frac{Q^{2} e^{2}}{16 \pi^{2}} p^{2} \,, \quad 
i {\cal M}_{2;(c); {\rm c.t.}} = 6 i \frac{Q^{2} e^{2}}{16 \pi^{2}} (k \cdot q) \,.
}
By summing them up, we obtain
\eqs{
i {\cal M}_{2; {\rm c.t.}} = - 3 i \frac{Q^{2} e^{2}}{16 \pi^{2}} (k^{2} + q^{2}) \,.
}
Note that out-going $\phi^{*} \phi$ are not necessarily on-shell.
Second, we evaluate \cref{fig:BPHZ_eom} in \cref{sec:oneloop_calc}:
\eqs{
i {\cal M}_{3;2; {\rm c.t.}}^{\mu} (p, q, k) = (i Q e {\tilde \mu}^{\epsilon / 2}) 6 \frac{Q^{2} e^{2}}{16 \pi^{2}} k^{\mu} \,, \quad i {\cal M}_{3;3; {\rm c.t.}}^{\mu} (p, q, k) = (i Q e {\tilde \mu}^{\epsilon / 2}) (-6) \frac{Q^{2} e^{2}}{16 \pi^{2}} q^{\mu} \,.
}
By summing them up, we obtain
\eqs{
i {\cal M}_{3; {\rm c.t.}}^{\mu} (p, q, k) = (i Q e {\tilde \mu}^{\epsilon / 2}) 6 \frac{Q^{2} e^{2}}{16 \pi^{2}} (k - q)^{\mu} \,.
}
The total contribution of the conterterms for diagrams with the insertion of $- (1/4) \epsilon F^{2}_{\mu \nu}$ are reproduced by insertion of the equation of motion:
\eqs{
3 \frac{Q^{2} e^{2}}{16 \pi^{2}} \left( \phi^{*} D^{2} \phi + D^{2} \phi^{*} \phi \right) \,.
}

The insertion of the equation of motion vanishes for the physical amplitude like scalaron decay $\sigma (p) \to \phi^{*} (q) \phi (k)$ considered in the main text.
[This is why we drop the (e.o.m.) term in \cref{eq:scalarclassicalT}.]
In conclusion, all the subdivergences, i.e., non-local divergences involved in two-loop diagrams of \cref{fig:twoloop} are canceled by the one-loop diagrams with the one-loop counterterms of \cref{fig:twoloop} in light of the BPHZ theorem~\cite{Bogoliubov:1957gp, Hepp:1966eg, Zimmermann:1969jj}.


\subsection{One-loop subdiagrams \label{sec:oneloop_calc}}

Here, we provide the computation of the one-loop subdiagrams which are illustrated in \cref{fig:BPHZ_eom,fig:oneloop_v3,fig:oneloop_v4}. 
We use divergent massless loop-integrals summarized in \cref{sec:loopfunc_YM}.
First, we evaluate diagrams in \cref{fig:BPHZ_eom} and determined counterterms.
\cref{fig:BPHZ_eom} (1) gives 
\eqs{
i {\cal M}_{3;1}^{\mu} (p, q, k) =& i \epsilon (i Q e {\tilde \mu}^{\epsilon / 2})^{3} \int \frac{d^{d} \ell}{(2 \pi)^{d}} \frac{- i g_{\beta \sigma}}{\ell^{2}} \left[ - \ell \cdot (\ell + p) g^{\rho \sigma} + \ell^{\rho} (\ell + p)^{\sigma} \right] \frac{- i g_{\alpha \rho}}{(\ell + p)^{2}} (- \ell - p + 2 q)^{\alpha} \\
& \times \frac{i}{(\ell + p - q)^{2}} (- 2 \ell - k - p + q)^{\mu} \frac{i}{(\ell + k)^{2}} (- \ell - 2k)^{\beta} \\
=& (i Q e {\tilde \mu}^{\epsilon / 2}) 4 i \epsilon Q^{2} e^{2} {\tilde \mu}^{\epsilon} \int \frac{d^{d} \ell}{(2 \pi)^{d}} \frac{(k \cdot q) \ell^{2} - (\ell \cdot k) (\ell \cdot q) + (k \cdot q) (\ell \cdot p) - (k \cdot p) (\ell \cdot q)}{[\ell^{2}][(\ell + p)^{2}][(\ell + p - q)^{2}][(\ell + k)^{2}]} \\
& \times (2 \ell + k + p - q)^{\mu}  \,.
}
Here, $p$ is an incoming momentum, while $q$ and $k$ are out-going momenta.
In the second equality, we use 
\eqs{
& g_{\beta \sigma} g_{\alpha \rho} \left[ - \ell \cdot (\ell + p) g^{\rho \sigma} + \ell^{\rho} (\ell + p)^{\sigma} \right] (\ell + p - 2q)^{\alpha} (\ell + 2k)^{\beta} \\
& = 4 \left[ (k \cdot q) \ell^{2} - (\ell \cdot k) (\ell \cdot q) + (k \cdot q) (\ell \cdot p) - (k \cdot p) (\ell \cdot q) \right] \,.
}
Thus, $i {\cal M}_{3;1}^{\mu} (p, q, k)$ is finite.

\cref{fig:BPHZ_eom} (2) gives 
\eqs{
i {\cal M}_{3;2}^{\mu} (p, q, k) =& i \epsilon (i Q e {\tilde \mu}^{\epsilon / 2}) (2 i Q^{2} e^{2} {\tilde \mu}^{\epsilon}) \int \frac{d^{d} \ell}{(2 \pi)^{d}} \frac{- i g_{\beta \sigma}}{\ell^{2}} \left[ - \ell \cdot (\ell + p) g^{\rho \sigma} + \ell^{\rho} (\ell + p)^{\sigma} \right] \frac{- i g_{\alpha \rho}}{(\ell + p)^{2}} \\
& \times g^{\alpha \mu} \frac{i}{(\ell + k)^{2}} (- \ell - 2k)^{\beta} \\
=& (i Q e {\tilde \mu}^{\epsilon / 2}) 4 i \epsilon Q^{2} e^{2} {\tilde \mu}^{\epsilon} \int \frac{d^{d} \ell}{(2 \pi)^{d}} \frac{k^{\mu} \ell^{2} - \ell^{\mu} (\ell \cdot k) + k^{\mu} (\ell \cdot p) - \ell^{\mu} (k \cdot p)}{[\ell^{2}][(\ell + p)^{2}][(\ell + k)^{2}]} \,.
}
In the second equality, we use 
\eqs{
& g_{\beta \sigma} g_{\alpha \rho} \left[ - \ell \cdot (\ell + p) g^{\rho \sigma} + \ell^{\rho} (\ell + p)^{\sigma} \right] g^{\alpha \mu} (\ell + 2k)^{\beta} \\
& = - 2 \left[ k^{\mu} \ell^{2} - \ell^{\mu} (\ell \cdot k) + k^{\mu} (\ell \cdot p) - \ell^{\mu} (k \cdot p)  \right] \,.
}
We use \cref{eq:loop-3} with the replacement of $k_{2} \to q$ and $\Delta_{xy} = - xy (1 - xy) k^{2} + 2 xy (1 - y) k \cdot p - y (1 - y) p^{2}$,
\eqs{
\left. i {\cal M}_{3;2}^{\mu} \right.|_{\rm div} (p, q, k) = (i Q e {\tilde \mu}^{\epsilon / 2}) (- 2) \frac{Q^{2} e^{2}}{16 \pi^{2}} \epsilon e^{\gamma_{E} (\epsilon / 2)} \Gamma(\epsilon / 2) (d - 1) k^{\mu} 
= (i Q e {\tilde \mu}^{\epsilon / 2}) (- 6) \frac{Q^{2} e^{2}}{16 \pi^{2}} k^{\mu} \,.
}

\cref{fig:BPHZ_eom} (3) gives 
\eqs{
i {\cal M}_{3;3}^{\mu} (p, q, k) =& i \epsilon (i Q e {\tilde \mu}^{\epsilon / 2}) (2 i Q^{2} e^{2} {\tilde \mu}^{\epsilon}) \int \frac{d^{d} \ell}{(2 \pi)^{d}} \frac{- i g_{\beta \sigma}}{\ell^{2}} \left[ - \ell \cdot (\ell + p) g^{\rho \sigma} + \ell^{\rho} (\ell + p)^{\sigma} \right] \frac{- i g_{\alpha \rho}}{(\ell + p)^{2}} \\
& \times (- \ell - p + 2q)^{\alpha} \frac{i}{(\ell + p - q)^{2}} g^{\beta \mu} \\
=& (i Q e {\tilde \mu}^{\epsilon / 2}) (- 4) i \epsilon Q^{2} e^{2} {\tilde \mu}^{\epsilon} \int \frac{d^{d} \ell}{(2 \pi)^{d}} \frac{q^{\mu} \ell^{2} - \ell^{\mu} (\ell \cdot q) + q^{\mu} (\ell \cdot p) - p^{\mu} (\ell \cdot q)}{[\ell^{2}][(\ell + p)^{2}][(\ell + p - q)^{2}]} \,.
}
In the second equality, we use 
\eqs{
& g_{\beta \sigma} g_{\alpha \rho} \left[ - \ell \cdot (\ell + p) g^{\rho \sigma} + \ell^{\rho} (\ell + p)^{\sigma} \right] (- \ell - p + 2q)^{\alpha} g^{\beta \mu} \\
& = 2 \left[ q^{\mu} \ell^{2} - \ell^{\mu} (\ell \cdot q) + q^{\mu} (\ell \cdot p) - p^{\mu} (\ell \cdot q)  \right] \,.
}
We use \cref{eq:loop-3} with the replacement of $k_{2} \to q$ and $\Delta_{xy} = - (1 - x) y (1 - y + x y) p^{2} + 2 x (1 - x) y^{2} p \cdot q - x y (1 - x y) q^{2}$,
\eqs{
\left. i {\cal M}_{3;3}^{\mu} \right.|_{\rm div} (p, q, k) = (i Q e {\tilde \mu}^{\epsilon / 2}) 2 \frac{Q^{2} e^{2}}{16 \pi^{2}} \epsilon e^{\gamma_{E} (\epsilon / 2)} \Gamma(\epsilon / 2) (d - 1) q^{\mu} 
= (i Q e {\tilde \mu}^{\epsilon / 2}) 6 \frac{Q^{2} e^{2}}{16 \pi^{2}} q^{\mu} \,.
}

Second, we compute the one-loop corrections to the $\phi^{*} (q) \phi (p) A_{\mu}$ vertex as shown in \cref{fig:oneloop_v3}.
Here $p$ is an incoming momentum, while $q$ is out-going.
\cref{fig:oneloop_v3} (1) gives
\eqs{
i \Gamma_{3;1}^{\mu} (p, q) =& (i Q e {\tilde \mu}^{\epsilon / 2})^{3} \int \frac{d^{d} \ell}{(2 \pi)^{d}} \frac{- i g_{\rho \sigma}}{\ell^{2}} (\ell + 2 p)^{\rho} \frac{i}{(\ell + p)^{2}} (2 \ell + p + q)^{\mu} \frac{i}{(\ell + q)^{2}} (\ell + 2 q)^{\sigma} \\
=& (i Q e {\tilde \mu}^{\epsilon / 2}) (- 1) i Q^{2} e^{2} {\tilde \mu}^{\epsilon} \int \frac{d^{d} \ell}{(2 \pi)^{d}} \frac{2 \ell^{\mu} \ell^{2} + 4 \ell^{\mu} \ell \cdot (p + q) + (p + q)^{\mu} \ell^{2} +\cdots}{[\ell^{2}] [(\ell + p)^{2}] [(\ell + q)^{2}]} \,.
}
In the second equality, we keep only the divergent part in the numerator, $(2 \ell + p + q)^{\mu} (\ell + 2 p) \cdot (\ell + 2 q)$.
We use \cref{eq:loop-3} with the replacement of $k_{2} \to q$, $q_{xy} = (1 - y) p + x y q$, and $\Delta_{xy} = - y (1 - y) p^{2} + 2 xy (1 - y) p \cdot q - xy (1 - xy) q^{2}$,
\eqs{
\left. i \Gamma_{3;1}^{\mu} \right|_{\rm div} (p, q) =& (i Q e {\tilde \mu}^{\epsilon / 2}) \frac{1}{2} \frac{Q^{2} e^{2}}{16 \pi^{2}} e^{\gamma_{E} (\epsilon / 2)} \Gamma(\epsilon / 2) \int_{0}^{1} dx dy y \left[ - 2 (d + 2) q_{xy}^{\mu} + 4 (p + q)^{\mu} + d (p + q)^{\mu} \right] \left( \frac{\mu^{2}}{\Delta_{xy}} \right)^{\epsilon / 2} \\
=& (i Q e {\tilde \mu}^{\epsilon / 2}) 2 (p + q)^{\mu} \frac{Q^{2} e^{2}}{16 \pi^{2}} \frac{1}{\epsilon}  \,.
}
In the second equality, we keep only the leading term.

\cref{fig:oneloop_v3} (2) and (2') give
\eqs{
i \Gamma_{3;2}^{\mu} (p, q) =& (i Q e {\tilde \mu}^{\epsilon / 2}) (2 i Q^{2} e^{2} {\tilde \mu}^{\epsilon}) \int \frac{d^{d} \ell}{(2 \pi)^{d}} \frac{- i g_{\rho \sigma}}{\ell^{2}} g^{\mu \sigma} \frac{i}{(\ell + p)^{2}} (\ell + 2 p)^{\rho} \,, \\
i \Gamma_{3;2'}^{\mu} (p, q) =& (i Q e {\tilde \mu}^{\epsilon / 2}) (2 i Q^{2} e^{2} {\tilde \mu}^{\epsilon}) \int \frac{d^{d} \ell}{(2 \pi)^{d}} \frac{- i g_{\rho \sigma}}{\ell^{2}} g^{\mu \sigma} \frac{i}{(\ell + q)^{2}} (\ell + 2 q)^{\rho} 
= i \Gamma_{3;2}^{\mu} (q, p) \,.
}
We use \cref{eq:loop-2} with the replacement of $q_{x} = xp$ and $\Delta_{x} = - x (1 - x) p^{2}$ for $i \Gamma_{3;2}^{\mu} (p, q)$ and that of $q_{x} = x q$ and $\Delta_{x} = - x (1 - x) q^{2}$ for $i \Gamma_{3;2'}^{\mu} (p, q)$,
\eqs{
\left. i \Gamma_{3;2}^{\mu} \right|_{\rm div} (p, q) =& (i Q e {\tilde \mu}^{\epsilon / 2}) (- 2) \frac{Q^{2} e^{2}}{16 \pi^{2}} e^{\gamma_{E} (\epsilon / 2)} \Gamma(\epsilon / 2) \int_{0}^{1} dx \left[ - q_{x} + 2 p^{\mu} \right] \left( \frac{\mu^{2}}{\Delta_{x}} \right)^{\epsilon / 2} \\
=& (i Q e {\tilde \mu}^{\epsilon / 2}) (-6) p^{\mu} \frac{Q^{2} e^{2}}{16 \pi^{2}} \frac{1}{\epsilon} \,, \\
\left. i \Gamma_{3;2'}^{\mu} \right|_{\rm div} (p, q) = & (i Q e {\tilde \mu}^{\epsilon / 2}) (-6) q^{\mu} \frac{Q^{2} e^{2}}{16 \pi^{2}} \frac{1}{\epsilon} \,.
}
Here, we keep only the leading terms.


Finally, we compute the one-loop corrections to the $\phi^{*} (q) \phi (p) A_{\mu} (k - p) A_{\nu}$ vertex as shown in \cref{fig:oneloop_v4}.
Here $p$ and $k - p$ are incoming momenta, while $q$ is out-going.
For \cref{fig:oneloop_v4} (2)-(5), there are contributions with the external photons being exchanged.
Since we are interested in the divergent part, we take them into account by multiplying $2$ in the end of calculations.
\cref{fig:oneloop_v4} (1) gives
\eqs{
i \Gamma^{\mu \nu}_{4;1} =& (2 i Q^{2} e^{2} {\tilde \mu}^{\epsilon}) (i Q e {\tilde \mu}^{\epsilon / 2})^{2} \int \frac{d^{d} \ell}{(2 \pi)^{d}} \frac{- i g_{\rho \sigma}}{\ell^{2}} (\ell + 2 p)^{\rho} \frac{i}{(\ell + p)^{2}} g^{\mu \nu} \frac{i}{(\ell + q)^{2}} (\ell + 2 q)^{\sigma} \\
=& (2 i Q^{2} e^{2} {\tilde \mu}^{\epsilon}) (- 1) g^{\mu \nu} i Q^{2} e^{2} {\tilde \mu}^{\epsilon} \int \frac{d^{d} \ell}{(2 \pi)^{d}} \frac{\ell^{2} + \cdots}{[\ell^{2}] [(\ell + p)^{2}] [(\ell + q)^{2}]} \,.
}
In the second equality, we keep only the divergent part in the numerator, $(\ell + 2 p) \cdot (\ell + 2 q)$.
We use \cref{eq:loop-3} with the replacement of $k_{2} \to q$ and $\Delta_{xy} = - y (1 - y) p^{2} + 2 xy (1 - y) p \cdot q - xy (1 - xy) q^{2}$,
\eqs{
\left. i \Gamma_{4;1}^{\mu \nu} \right|_{\rm div} =& (2 i Q^{2} e^{2} \mu^{\epsilon}) \frac{d}{4} g^{\mu \nu} \frac{Q^{2} e^{2}}{16 \pi^{2}} e^{\gamma_{E} (\epsilon / 2)} \Gamma(\epsilon / 2) \int_{0}^{1} dx dy \, 2 y \left( \frac{\mu^{2}}{\Delta_{xy}} \right)^{\epsilon / 2} \\
=& (2 i Q^{2} e^{2} \mu^{\epsilon}) 2 g^{\mu \nu} \frac{Q^{2} e^{2}}{16 \pi^{2}} \frac{1}{\epsilon}  \,.
}
In the second equality, we keep only the leading term.

\cref{fig:oneloop_v4} (2) gives
\eqs{
i \Gamma^{\mu \nu}_{4;2} =& (2 i Q^{2} e^{2} {\tilde \mu}^{\epsilon}) (i Q e {\tilde \mu}^{\epsilon / 2})^{2} \int \frac{d^{d} \ell}{(2 \pi)^{d}} \frac{- i g_{\rho \sigma}}{\ell^{2}} g^{\mu \rho} \frac{i}{(\ell + k)^{2}} (2 \ell + k + q)^{\nu} \frac{i}{(\ell + q)^{2}} (\ell + 2 q)^{\sigma} \\
=& (2 i Q^{2} e^{2} {\tilde \mu}^{\epsilon}) (- 1) i Q^{2} e^{2} {\tilde \mu}^{\epsilon} \int \frac{d^{d} \ell}{(2 \pi)^{d}} \frac{2 \ell^{\mu} \ell^{\nu} + \cdots}{[\ell^{2}] [(\ell + k)^{2}] [(\ell + q)^{2}]} \,.
}
In the second equality, we keep only the divergent part in the numerator, $(\ell + 2 q)^{\mu} (2 \ell + k + q)^{\nu}$.
We use \cref{eq:loop-3}, where $k_{2} \to k$ and $p \to q$ and $\Delta_{xy} = - xy (1 - xy) k^{2} + 2 xy (1 - y) k \cdot q - y (1 - y) q^{2}$,
\eqs{
\left. i \Gamma^{\mu \nu}_{4;2} \right|_{\rm div} =& (2 i Q^{2} e^{2} {\tilde \mu}^{\epsilon}) \frac{1}{2} \frac{Q^{2} e^{2}}{16 \pi^{2}} e^{\gamma_{E} (\epsilon / 2)} \Gamma(\epsilon / 2) \int_{0}^{1} dx dy \, 2 y g^{\mu \nu} \left( \frac{\mu^{2}}{\Delta_{xy}} \right)^{\epsilon / 2} \\
=& (2 i Q^{2} e^{2} \mu^{\epsilon}) g^{\mu \nu} \frac{Q^{2} e^{2}}{16 \pi^{2}} \frac{1}{\epsilon}  \,.
}
In the second equality, we keep only the leading term.
This contribution should be doubled to take into account the external-photon exchange.

\cref{fig:oneloop_v4} (3) gives
\eqs{
i \Gamma^{\mu \nu}_{4;3} =& (2 i Q^{2} e^{2} {\tilde \mu}^{\epsilon}) (i Q e {\tilde \mu}^{\epsilon / 2})^{2} \int \frac{d^{d} \ell}{(2 \pi)^{d}} \frac{- i g_{\rho \sigma}}{\ell^{2}} (\ell + 2 p)^{\rho} \frac{i}{(\ell + p)^{2}} (2 \ell + k + p)^{\mu} \frac{i}{(\ell + k)^{2}} g^{\nu \sigma}  \\
=& (2 i Q^{2} e^{2} \mu^{\epsilon}) (- 1) i Q^{2} e^{2} {\tilde \mu}^{\epsilon} \int \frac{d^{d} \ell}{(2 \pi)^{d}} \frac{2 \ell^{\mu} \ell^{\nu} + \cdots}{[\ell^{2}] [(\ell + k)^{2}] [(\ell + p)^{2}]} \,.
}
In the second equality, we keep only the divergent part in the numerator, $(2 \ell + k + p)^{\mu} (\ell + 2 p)^{\nu}$.
We use \cref{eq:loop-3} with the replacement of $k_{2} \to k$ and $\Delta_{xy} = - xy (1 - xy) k^{2} + 2 xy (1 - y) k \cdot p - y (1 - y) p^{2}$,
\eqs{
\left. i \Gamma^{\mu \nu}_{4;3} \right|_{\rm div} =& (2 i Q^{2} e^{2} {\tilde \mu}^{\epsilon}) \frac{1}{2} \frac{Q^{2} e^{2}}{16 \pi^{2}} e^{\gamma_{E} (\epsilon / 2)} \Gamma(\epsilon / 2) \int_{0}^{1} dx dy \, 2 y g^{\mu \nu} \left( \frac{\mu^{2}}{\Delta_{xy}} \right)^{\epsilon / 2} \\
=& (2 i Q^{2} e^{2} {\tilde \mu}^{\epsilon}) g^{\mu \nu} \frac{Q^{2} e^{2}}{16 \pi^{2}} \frac{1}{\epsilon}  \,.
}
In the second equality, we keep only the leading term.
This contribution should be doubled to take into account the external-photon exchange.

\cref{fig:oneloop_v4} (4) gives
\eqs{
i \Gamma^{\mu \nu}_{4;4} =& (i Q e {\tilde \mu}^{\epsilon / 2})^{4} \int \frac{d^{d} \ell}{(2 \pi)^{d}} \frac{- i g_{\rho \sigma}}{\ell^{2}} (\ell + 2 p)^{\rho} \frac{i}{(\ell + p)^{2}} (2 \ell + k + p)^{\mu} \frac{i}{(\ell + k)^{2}} (2 \ell + k + q)^{\nu} \frac{i}{(\ell + q)^{2}} (\ell + 2 q)^{\sigma} \\
=& (2 i Q^{2} e^{2} {\tilde \mu}^{\epsilon}) \frac{1}{2} g_{\rho \sigma} i Q^{2} e^{2} {\tilde \mu}^{\epsilon} \int \frac{d^{d} \ell}{(2 \pi)^{d}} \frac{4 \ell^{\mu} \ell^{\nu} \ell^{\rho} \ell^{\sigma} + \cdots}{[\ell^{2}] [(\ell + k)^{2}] [(\ell + p)^{2}] [(\ell + q)^{2}]} \,.
}
In the second equality, we keep only the divergent part in the numerator, $(2 \ell + k + p)^{\mu} (2 \ell + k + q)^{\nu} (\ell + 2 p)^{\rho} (\ell + 2 q)^{\sigma}$.
We use \cref{eq:loop-3},
\eqs{
\left. i \Gamma^{\mu \nu}_{4;4} \right|_{\rm div} =& (2 i Q^{2} e^{2} {\tilde \mu}^{\epsilon}) (-1) \frac{d+2}{8} \frac{Q^{2} e^{2}}{16 \pi^{2}} e^{\gamma_{E} (\epsilon / 2)} \Gamma(\epsilon / 2) \int_{0}^{1} dx dy dz \, 4 y z^{2} g^{\mu \nu} \left( \frac{\mu^{2}}{\Delta_{xy}} \right)^{\epsilon / 2} \\
=& (2 i Q^{2} e^{2} \mu^{\epsilon}) (- 1) g^{\mu \nu} \frac{Q^{2} e^{2}}{16 \pi^{2}} \frac{1}{\epsilon}  \,.
}
In the second equality, we keep only the leading term.
This contribution should be doubled to take into account the external-photon exchange.

\cref{fig:oneloop_v4} (5) gives
\eqs{
i \Gamma^{\mu \nu}_{4;5} =& (2 i Q^{2} e^{2} {\tilde \mu}^{\epsilon})^{2} \int \frac{d^{d} \ell}{(2 \pi)^{d}} \frac{- i g_{\rho \sigma}}{\ell^{2}} g^{\mu \rho} \frac{i}{(\ell + k)^{2}} g^{\nu \sigma}  \\
=& (2 i Q^{2} e^{2} {\tilde \mu}^{\epsilon}) 2 g^{\mu \nu} i Q^{2} e^{2} {\tilde \mu}^{\epsilon} \int \frac{d^{d} \ell}{(2 \pi)^{d}} \frac{1}{[\ell^{2}] [(\ell + k)^{2}]} \,.
}
We use \cref{eq:loop-2} with the replacement of $p \to k$ and $\Delta_{x} = - x (1 - x) k^{2}$
\eqs{
\left. i \Gamma^{\mu \nu}_{4;5} \right|_{\rm div} =& (2 i Q^{2} e^{2} {\tilde \mu}^{\epsilon}) (-2) g^{\mu \nu} \frac{Q^{2} e^{2}}{16 \pi^{2}} e^{\gamma_{E} (\epsilon / 2)} \Gamma(\epsilon / 2) \int_{0}^{1} dx \left( \frac{\mu^{2}}{\Delta_{x}} \right)^{\epsilon / 2} \\
=& (2 i Q^{2} e^{2} \mu^{\epsilon}) (- 4) g^{\mu \nu} \frac{Q^{2} e^{2}}{16 \pi^{2}} \frac{1}{\epsilon}  \,.
}
In the second equality, we keep only the leading term.
This contribution should be doubled to take into account the external-photon exchange.

\section{$\beta_{e}$ in the QCD \label{sec:YM}}

We consider the QCD whose action is given by
\eqs{
S_{\rm mat} = & \int d^{4} x \sqrt{- g} \left( - \frac{1}{4} g^{\mu \lambda} g^{\nu \kappa} F^{a}_{0 \mu \nu} F^{a}_{0 \lambda \kappa} \right) + S_{\rm fix} \,.
}
where we omit the other fields (e.g., quarks) for simplicity.
The field strength is
\eqs{
F^{a}_{0 \mu \nu} = \nabla_{\mu} A^{a}_{0 \nu} - \nabla_{\nu} A^{a}_{0 \mu} + e_{0} f^{a b c} A^{b}_{0 \mu} A^{c}_{0 \nu} \,,
}
with the structure constant $f^{a b c}$.
We omit the gauge fixing term $S_{\rm fix}$ for $T^{\mu \nu}$ as discussed in Ref.~\cite{Kamada:2019pmx}.
The $d$-dimensional flat-spacetime energy-momentum tensor is given by
\eqs{
T_{\mu \nu} = - g^{\lambda \kappa} F^{a}_{0 \mu \lambda} F^{a}_{0 \nu \kappa} + \frac{1}{4} g_{\mu \nu} g^{\lambda \rho} g^{\kappa \sigma} F^{a}_{0 \lambda \kappa} F^{a}_{0 \rho \sigma} \,.
}
Taking the trace, one finds
\eqs{
\label{eq:Tmumu_YM}
T^{\mu}_{~ \mu} &= - \frac{1}{4} \epsilon F_{0 \mu \nu}^{2} \\
& = - \frac{1}{2} \epsilon (\partial^{\mu} A^{a \nu} - \partial^{\nu} A^{a \mu}) \partial_{\mu} A^{a}_{\nu} - \epsilon {\tilde \mu}^{\epsilon / 2} e f^{a b c} A^{a \mu} A^{b \nu} \partial_{\mu} A^{c}_{\nu} - \frac{1}{4} \epsilon {\tilde \mu}^{\epsilon} e^{2}  f^{a b e} f^{c d e} A^{a}_{\mu} A^{b}_{\nu} A^{c \mu} A^{d \nu} \\
& \quad - \frac{1}{2} \epsilon (Z_{A} - 1) (\partial^{\mu} A^{a \nu} - \partial^{\nu} A^{a \mu}) \partial_{\mu} A^{a}_{\nu} - \epsilon (Z_{e} - 1) {\tilde \mu}^{\epsilon / 2} e f^{a b c} A^{a \mu} A^{b \nu} \partial_{\mu} A^{c}_{\nu} \\ 
& \quad - \frac{1}{4} \epsilon (Z_{e^{2}} - 1) {\tilde \mu}^{\epsilon} e^{2}  f^{a b e} f^{c d e} A^{a}_{\mu} A^{b}_{\nu} A^{c \mu} A^{d \nu} \,.
}
Here we introduce the multiplicative renormalization:
\eqs{
A_{0 \mu} = Z_{A}^{1/2} A_{\mu} \,,
}
and
\eqs{
Z_{A}^{3/2} e_{0} = Z_{e} {\tilde \mu}^{\epsilon / 2} e \,, \quad Z_{A}^{2} e_{0}^{2} = Z_{e^{2}} {\tilde \mu}^{\epsilon} e^{2} \quad ({\rm i.e.}, Z_{A} Z_{e^{2}}= Z_{e}^{2}) \,.
}
We summarize divergent one-loop integrals in \cref{sec:loopfunc_YM}.
We use Feynman-'t Hooft gauge $(\xi^\mathrm{gf} = 1)$ in the loop calculations.

The scalar and fermion contributions to the wavefunction renormalization is similar to those in the QED (see Ref.~\cite{Kamada:2019pmx}):
\eqs{
Z_{A} - 1 \supset \left( - \frac{4}{3} T(F) - \frac{2}{3} T(S) \right)\frac{e^{2}}{16 \pi^{2}} \frac{1}{\epsilon} \,,
}
except for that $Q^{2}$ is replaced by $T (S)$ and $T(F)$ denoting one-halves of the Dynkin index of the representation for the scalar and fermion (Weyl) fields, respectively.
There are additional contributions from gauge-boson and ghost loops.
The divergent part takes the form of [see \cref{eq:loop-2}]
\eqs{
\left. i \Gamma^{(a \mu)(b \nu)} \right|_{\rm div} (p) =& - e^{2} T({\rm Ad}) \delta^{a b} \left[ \frac{i}{16 \pi^{2}} e^{\gamma_{E} (\epsilon / 2)} \left( - \frac{1}{2} \right) \Gamma(- 1 +  \epsilon / 2) \int_{0}^{1} dx \, h (x) g^{\mu \nu} \Delta_{x} \left( \frac{\mu^{2}}{\Delta_{x}} \right)^{\epsilon / 2} \right. \\
& \left.  + \frac{i}{16 \pi^{2}} e^{\gamma_{E} (\epsilon / 2)} \Gamma(\epsilon / 2) \int_{0}^{1} dx \, \left[ f (x) p^{2} g^{\mu \nu} + g (x) p^{\mu} p^{\nu} \right] \left( \frac{\mu^{2}}{\Delta_{x}} \right)^{\epsilon / 2} \right] \,,
}
where $\Delta_{x} = - x (1 -x) p^{2}$, and $T (\mathrm{Ad})$ denotes the one-half of the Dynkin index of the adjoint representation for the gauge field.
We determine $f (x)$, $g (x)$, and $h (x)$ below.

\begin{figure}
	\centering
	\includegraphics[width=0.8\linewidth]{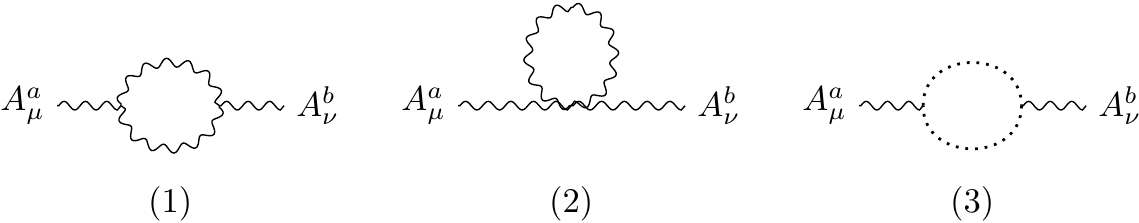}
	\caption{
		One-loop self-energy diagrams for the QCD.
		First-two diagrams are contributions from the self-interactions of gauge bosons, while the last is a contribution from the ghost interaction.
	}
	\label{fig:YM_self}
\end{figure}

\cref{fig:YM_self} shows contributions from gauge-boson and ghost loops.
The diagram (1) in \cref{fig:YM_self} provides
\eqs{
i \Gamma_{1}^{(a \mu)(b \nu)} (p) =& (e \mu^{\epsilon / 2})^{2} \frac{1}{2} \int \frac{d^{d} \ell}{(2 \pi)^{d}} \frac{- i g_{\rho \alpha} \delta^{c e}}{\ell^{2}} f^{a c d} V_{e}^{\mu \rho \sigma} (p, \ell, - \ell - p) \frac{- i g_{\sigma \beta} \delta^{d f}}{(\ell + p)^{2}} f^{b e f}  V_{e}^{\nu \alpha \beta} (- p, - \ell, \ell + p) \\
=& - e^{2} T({\rm Ad}) \delta^{a b} \mu^{\epsilon}\int \frac{d^{d} \ell}{(2 \pi)^{d}} \frac{1}{[\ell^{2}] [(\ell + p)^{2}]} \left[ - g^{\mu \nu} \ell^{2} - (2d - 3) \ell^{\mu} \ell^{\nu} \right. \\ 
& \left. - g^{\mu \nu} \ell \cdot p - \frac{1}{2} (2d - 3) \ell^{\mu} p^{\nu} - \frac{1}{2} (2d - 3) p^{\mu} \ell^{\nu} - \frac{5}{2} g^{\mu \nu} p^{2} + \frac{1}{2} (6 - d) p^{\mu} p^{\nu} \right] \,.
}
In the first equality, we take into account the symmetric factor of $1/2$.
$V_{e}^{\mu \nu \rho} (p, q, k) = g^{\mu \nu} (p - q)^{\rho} + g^{\nu \rho} (q - k)^{\mu} + g^{\rho \mu} (k - p)^{\nu}$ is the vertex factor with $p$, $q$, and $k$ being incoming momenta of $A^{a \mu}$, $A^{b \nu}$, and $A^{c \rho}$, respectively.
In the second equality, we use $\delta^{c e} \delta^{d f} f^{a c d} f^{b e f} = T({\rm Ad}) \delta^{a b}$ and
\eqs{
& g_{\rho \alpha} g_{\sigma \beta} V_{e}^{\mu \rho \sigma} (p, \ell, - \ell - p) V_{e}^{\nu \alpha \beta} (- p, - \ell, \ell + p) \\
& = - 2 g^{\mu \nu} \ell^{2} - 2 (2d - 3) \ell^{\mu} \ell^{\nu} - 2 g^{\mu \nu}\ell \cdot p  - (2d - 3) \ell^{\mu} p^{\nu} - (2d - 3) p^{\mu} \ell^{\nu} - 5 g^{\mu \nu} p^{2}+ (6 - d) p^{\mu} p^{\nu} \,.
}
It contributes to the divergent part as [see \cref{eq:loop-2}]
\eqs{
& h_{1} (x) = - 3 (d - 1) \,, \quad f_{1} (x) = - \frac{5}{2} + x - x^{2} \,, \\
& g_{1} (x) = \frac{1}{2} (6 - d) + (2d - 3) x - (2d - 3) x^{2} \,.
}

The diagram (2) in \cref{fig:YM_self} provides
\eqs{
i \Gamma_{2}^{(a \mu)(b \nu)} (p) =& - i e^{2} {\tilde \mu}^{\epsilon} \frac{1}{2} \int \frac{d^{d} \ell}{(2 \pi)^{d}} \frac{- i g_{\rho \sigma} \delta^{c d}}{\ell^{2}} V_{e^{2}}^{(a \mu) (b \nu) (c \rho) (d \sigma)} \\
=& - e^{2} T({\rm Ad}) \delta^{a b} {\tilde \mu}^{\epsilon} \int \frac{d^{d} \ell}{(2 \pi)^{d}} \frac{(d - 1) (\ell^{2} + 2 \ell \cdot p  + p^{2}) g^{\mu \nu}}{[\ell^{2}] [(\ell + p)^{2}]}  \,.
}
In the first equality, we take into account the symmetric factor of $1/2$.
$V_{e^{2}}^{(a \mu) (b \nu) (c \rho) (d \sigma)} = f^{a b e} f^{c d e} (g^{\mu \rho} g^{\nu \sigma} - g^{\mu \sigma} g^{\nu \rho}) + f^{a c e} f^{b d e} (g^{\mu \nu} g^{\rho \sigma} - g^{\mu \sigma} g^{\nu \rho}) + f^{a d e} f^{b c e} (g^{\mu \nu} g^{\rho \sigma} - g^{\mu \rho} g^{\nu \sigma})$ is the vertex factor for $A^{a \mu}$, $A^{b \nu}$, $A^{c \rho}$, and $A^{d \sigma}$.
In the second equality, we use $g_{\rho \sigma} \delta^{c d} V_{e^{2}}^{(a \mu) (b \nu) (c \rho) (d \sigma)}  = 2 (d - 1) T({\rm Ad}) g^{\mu \nu}$ and multiply $(\ell + p)^{2}$ in both the numerator and denominator of the integrand.
It contributes to the divergent part as [see \cref{eq:loop-2}]
\eqs{
h_{2} (x) = d (d - 1) \,, \quad  f_{2} (x) = (d - 1) - 2 (d - 1) x + (d - 1) x^{2} \,, \quad g_{2} (x) = 0 \,.
}

The diagram (3) in \cref{fig:YM_self} provides 
\eqs{
i \Gamma_{3}^{(a \mu)(b \nu)} (p) =& (- e \mu^{\epsilon / 2})^{2} (-1) {\tilde \mu}^{\epsilon} \int \frac{d^{d} \ell}{(2 \pi)^{d}} \frac{i \delta^{c e}}{\ell^{2}} f^{a c d} (\ell + p)^{\mu} \frac{i \delta^{d f}}{(\ell + p)^{2}} f^{b f e} \ell^{\nu} \\
=& - e^{2} T({\rm Ad}) \delta^{a b} {\tilde \mu}^{\epsilon} \int \frac{d^{d} \ell}{(2 \pi)^{d}} \frac{\ell^{\mu} \ell^{\nu} + p^{\mu} \ell^{\nu}}{[\ell^{2}] [(\ell + p)^{2}]}  \,.
}
In the first equality, we take into account the ghost statistics of $(-1)$.
In the second equality, we use $\delta^{c e} \delta^{d f} f^{a c d} f^{b f e} = - T({\rm Ad}) \delta^{a b}$.
It contributes to the divergent part as [see \cref{eq:loop-2}]
\eqs{
h_{3} (x) = 1 \,, \quad  f_{3} (x) = 0 \,, \quad g_{3} (x) = - x + x^{2} \,.
}

In summary,
\eqs{
& h (x) = (d - 2)^{2} \,, \quad f (x) = \frac{1}{2} (2 d - 7) - (2d - 3) x + (d - 2) x^{2} \,, \\
& \quad g(x) = \frac{1}{2} (6 - d) + 2 (d - 2) x - 2 (d - 2) x^{2} \,.
}
Noting that $(d - 2) (- 1 / 2) \Gamma (-1 + \epsilon / 2) = \Gamma (\epsilon / 2)$ (i.e., no quadratic divergence), we obtain the divergent part of
\eqs{
\left. i \Gamma^{(a \mu)(b \nu)} \right|_{\rm div} (p) =& - i \frac{e^{2}}{16 \pi^{2}} T({\rm Ad}) \delta^{a b} e^{\gamma_{E} (\epsilon / 2)} \Gamma(\epsilon / 2) \int_{0}^{1} dx \, \left[ F (x) p^{2} g^{\mu \nu} + G (x) p^{\mu} p^{\nu} \right] \left( \frac{\mu^{2}}{\Delta_{x}} \right)^{\epsilon / 2} \,,
}
where 
\eqs{
& F (x) = (1 - 3 x + 2 x^{2}) (d - 4) + \left( \frac{1}{2} - 7 x + 4 x^{2} \right) \,, \\
& G(x) = \left( - \frac{1}{2} + 2 x - 2 x^{2} \right) (d - 4) + (1 + 4 x - 4 x^{2}) \,, \\
& \int_{0}^{1} dx F (x) = - \int_{0}^{1} dx G (x) = \frac{1}{6} (d - 4) - \frac{5}{3} \,,
}
and thus,
\eqs{
\left( i \Gamma^{(a \mu)(b \nu)} \right)^{\rm pole}_{\rm of \, \epsilon} (p) =& i (p^{2} g^{\mu \nu} - p^{\mu} p^{\nu}) \delta^{a b} \frac{e^{2}}{16 \pi^{2}} \frac{10}{3} T({\rm Ad}) \frac{1}{\epsilon} \,.
}

As a result, the wavefunction renormalization in Feynman-'t Hooft gauge is given by
\eqs{
Z_{A} - 1 = \left( \frac{10}{3} T({\rm Ad}) - \frac{4}{3} T(F) - \frac{2}{3} T(S) \right) \frac{e^{2}}{16 \pi^{2}} \frac{1}{\epsilon} \,.
}
Meanwhile, the $\beta$ function of the gauge coupling is known to be
\eqs{
\label{eq:betae_YM}
\beta_{e} = - \left( \frac{11}{3} T({\rm Ad}) - \frac{2}{3} T(F) - \frac{1}{3} T(S) \right) \frac{e^{3}}{16 \pi^{2}} \,.
}
Thus, $\epsilon (Z_{A} - 1) = - 2 \beta_{e} / e$ does not hold.
Precisely speaking, it holds for the scalar and fermion contributions, but not for the gauge boson contribution.
We note that $Z_{A}$ is gauge dependent in the QCD, while $\beta_{e}$ is gauge independent.
On the other hand, since $T^{\mu}_{~ \mu}$ is gauge invariant, we expect that the invariant amplitude of scalaron decay into two gauge bosons is proportional to $\beta_{e}$ in the QCD as in the QED~\cite{Kamada:2019pmx}.
We directly check it in the following.
There, the self-coupling terms of $T^{\mu}_{~ \mu}$ [second and third terms of \cref{eq:Tmumu_YM}] become important.

We consider scalaron decay into two gauge bosons, $\sigma (p) \to A^{a}_{\mu} (k_{1}) A^{b} _{\nu} (k_{2})$ ($p$: incoming momentum; $k_{1}$ and $k_{2}$: outgoing momenta), at the one-loop level.
There are two types contributions from the insertion of $T^{\mu}_{~ \mu} \supset - (1/4) \epsilon F^{a 2}_{0 \mu \nu}$: 1) tree-level diagram from the insertion of $T^{\mu}_{~ \mu} \supset - (1/2) \epsilon (Z_{A} - 1) (\partial^{\mu} A^{a \nu} - \partial^{\nu} A^{a \mu}) \partial_{\mu} A^{a}_{\nu}$; and 2) one-loop diagrams from the insertion of 
\eqs{
T^{\mu}_{~ \mu} \supset - (1/2) \epsilon (\partial^{\mu} A^{a \nu} - \partial^{\nu} A^{a \mu}) \partial_{\mu} A^{a}_{\nu} - \epsilon  {\tilde \mu}^{\epsilon / 2} e f^{a b c} A^{a \mu} A^{b \nu} \partial_{\mu} A^{c}_{\nu} - (1/4) \epsilon  {\tilde \mu}^{\epsilon} e^{2}  f^{a b e} f^{c d e} A^{a}_{\mu} A^{b}_{\nu} A^{c \mu} A^{d \nu} \,.
}
The type-1) contribution is simply given by
\eqs{
\label{eq:YM_tree}
i {\cal M}_{\rm tree} = i \epsilon (Z_{A} - 1) (k_{1} \cdot k_{2} g^{\mu \nu} - k_{2}^{\mu} k_{1}^{\nu}) \epsilon_{1 \mu}^{*} \epsilon_{2 \nu}^{*} \delta^{a b} \,,
}
with polarization vectors of $k_{1} \cdot \epsilon_{1} = k_{2} \cdot \epsilon_{2} = 0$.
The divergent part of the type-2) contributions takes the form of [see \cref{eq:loop-3}]
\eqs{
\left. i {\cal M}_{\rm loop} \right|_{\rm div} =& - \epsilon e^{2} T({\rm Ad}) \delta^{a b} \epsilon_{1 \mu}^{*} \epsilon_{2 \nu}^{*}  \left[ \frac{i}{16 \pi^{2}} e^{\gamma_{E} (\epsilon / 2)} \left( - \frac{d + 2}{4} \right) \Gamma(- 1 +  \epsilon / 2) \int_{0}^{1} dx dy y \, h (x, y) g^{\mu \nu} \Delta_{xy} \left( \frac{\mu^{2}}{\Delta_{xy}} \right)^{\epsilon / 2} \right. \\
& \left. + \frac{1}{2} \frac{i}{16 \pi^{2}} e^{\gamma_{E} (\epsilon / 2)} \Gamma(\epsilon / 2) \int_{0}^{1} dx dy y \, \left[ f (x, y) k_{1} \cdot k_{2} g^{\mu \nu} + g (x, y) k_{2}^{\mu} k_{1}^{\nu} \right] \left( \frac{\mu^{2}}{\Delta_{xy}} \right)^{\epsilon / 2} \right] \,,
}
where $\Delta_{xy} = - 2 y (1 - x) (1 - y) k_{1} \cdot k_{2}$.
Here we use $k_{1} \cdot \epsilon_{1} = k_{2} \cdot \epsilon_{2} = 0$ and $k_{1}^{2} = k_{2}^{2} = 0$.
We determine $f (x, y)$, $g (x, y)$, and $h (x, y)$ below.

\begin{figure}
	\centering
	\includegraphics[width=0.8\linewidth]{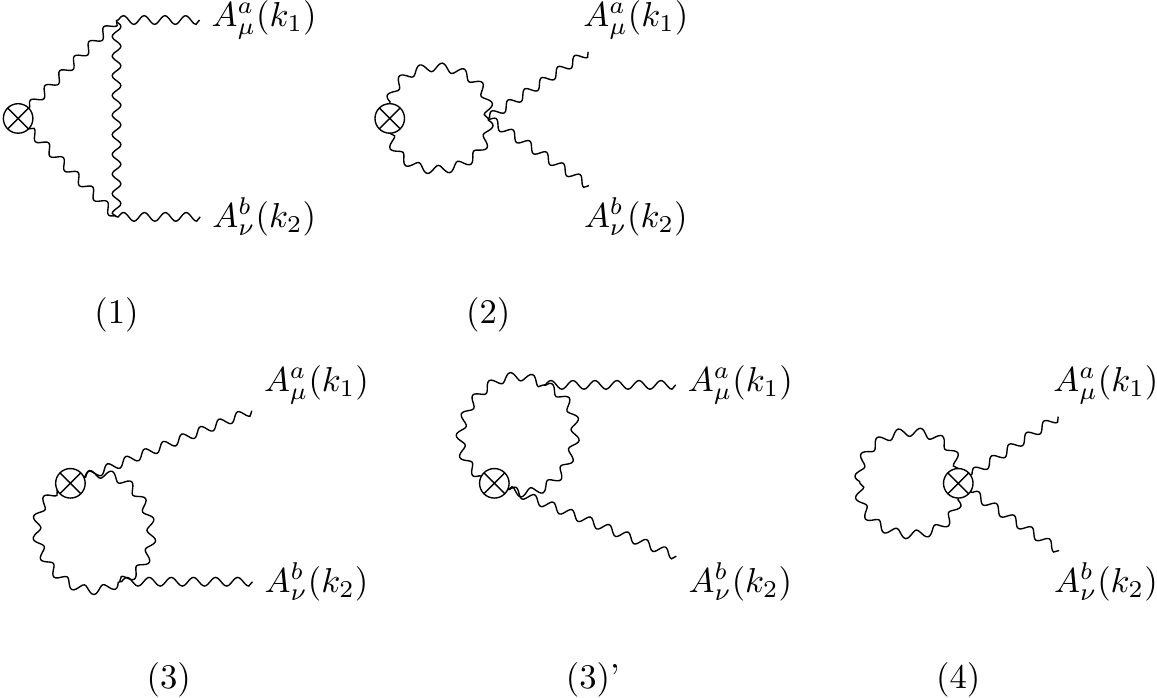}
	\caption{Type-2) one-loop diagrams for scalaron decay $\sigma (p) \to A^{a}_{\mu} (k_{1}) A^{b} _{\nu} (k_{2})$.
	Crossed dots denote insertion of the energy-momentum tensor, $- (1/2) \epsilon (\partial^{\mu} A^{a \nu} - \partial^{\nu} A^{a \mu}) \partial_{\mu} A^{a}_{\nu}$ (1, 2), $- \epsilon  {\tilde \mu}^{\epsilon / 2} e f^{a b c} A^{a \mu} A^{b \nu} \partial_{\mu} A^{c}_{\nu}$ (3, 4), and $- (1/4) \epsilon  {\tilde \mu}^{\epsilon} e^{2}  f^{a b e} f^{c d e} A^{a}_{\mu} A^{b}_{\nu} A^{c \mu} A^{d \nu}$ (5).}
	\label{fig:YM_loop}
\end{figure}

\cref{fig:YM_loop} show the type-2) one-loop diagrams.
\cref{fig:YM_loop} ($1$) and ($2$) are from the insertion of $T^{\mu}_{~ \mu} \supset - (1/2) \epsilon (\partial^{\mu} A^{a \nu} - \partial^{\nu} A^{a \mu}) \partial_{\mu} A^{a}_{\nu}$.
\cref{fig:YM_loop} ($1$) gives
\eqs{
i {\cal M}_{1} =& i \epsilon (e {\tilde \mu}^{\epsilon / 2})^{2} \int \frac{d^{d} \ell}{(2 \pi)^{d}} \frac{- i g_{\beta \sigma} \delta^{h e}}{\ell^{2}} \left[- \ell \cdot (\ell + p) g^{\alpha \beta} + \ell^{\alpha} (\ell + p)^{\beta} \right] \delta^{g h} \frac{- i g_{\alpha \rho} \delta^{g c}}{(\ell + p)^{2}}  \\ 
& \times \epsilon_{1 \mu}^{*}  f^{a c d} V_{e}^{\mu \rho \gamma} (- k_{1}, \ell + p, - \ell - k_{2}) \frac{- i g_{\delta \gamma} \delta^{f d}}{(\ell + k_{2})^{2}} \epsilon_{2 \nu}^{*} f^{b e f}  V_{e}^{\nu \sigma \delta} (- k_{2}, - \ell, \ell + k_{2}) \\
=& - \epsilon e^{2} T({\rm Ad}) \delta^{a b} \epsilon_{1 \mu}^{*} \epsilon_{2 \nu}^{*} {\tilde \mu}^{\epsilon} \int \frac{d^{d} \ell}{(2 \pi)^{d}} \frac{1}{[\ell^{2}] [(\ell + k_{2})^{2}] [(\ell + p)^{2}]} \left[ g^{\mu \nu} (\ell^{2})^{2}  + (4d - 5) \ell^{\mu} \ell^{\nu} \ell^{2} \right. \\
& \left. + g^{\mu \nu} \ell^{2} \ell \cdot (3 k_{1} + k_{2}) + 2 \ell^{\mu} \ell^{\nu} \ell \cdot \left[ (2d - 3) k_{1} + 2 (d - 1) k_{2} \right] - \ell^{\mu} k_{1}^{\nu} \ell^{2} + 4 (d - 1) k_{2}^{\mu} \ell^{\nu} \ell^{2} \right. \\
& \left. - 6 g^{\mu \nu} \ell^{2} k_{1} \cdot k_{2} + 6 g^{\mu \nu} (\ell \cdot k_{1}) \ell \cdot k_{2} + 10 \ell^{\mu} \ell^{\nu} k_{1} \cdot k_{2} - 8 \ell^{\mu} k_{1}^{\nu} \ell \cdot k_{2} + 4 (d - 3) k_{2}^{\mu} \ell^{\nu} \ell \cdot k_{1}  \right. \\
& \left. + 8 k_{2}^{\mu} k_{1}^{\nu} \ell^{2} + \cdots \right] \,.
}
In the second equality, we use $\delta^{h e} \delta^{g h} \delta^{g c} \delta^{f d} f^{a c d} f^{b e f} = T({\rm Ad}) \delta^{a b}$ and
\eqs{
& g_{\beta \sigma} g_{\alpha \rho} g_{\delta \gamma} \left[- \ell \cdot (\ell + p) g^{\alpha \beta} + \ell^{\alpha} (\ell + p)^{\beta} \right] V_{e}^{\mu \rho \gamma} (- k_{1}, \ell + p, - \ell - k_{2}) V_{e}^{\nu \sigma \delta} (- k_{2}, - \ell, \ell + k_{2}) \\
& = g^{\mu \nu} (\ell^{2})^{2}  + (4d - 5) \ell^{\mu} \ell^{\nu} \ell^{2} + g^{\mu \nu} \ell^{2} \ell \cdot (3 k_{1} + k_{2}) + 2 \ell^{\mu} \ell^{\nu} \ell \cdot \left[ (2d - 3) k_{1} + 2 (d - 1) k_{2} \right] \\
& \quad + \ell^{\mu} \left[ - k_{1} + (2d - 3) k_{2} \right]^{\nu} \ell^{2} + 2 (d - 1) (k_{1} + 2 k_{2})^{\mu} \ell^{\nu} \ell^{2} - g^{\mu \nu} \ell^{2} (6   k_{1} \cdot k_{2} + k_{2}^{2}) \\
& \quad + 2 g^{\mu \nu} (\ell \cdot k_{1}) \ell \cdot (k_{1} + 3 k_{2}) + \ell^{\mu} \ell^{\nu} (10 k_{1} \cdot k_{2} + 4 k_{1}^{2} + 5 k_{2}^{2}) - 2 \ell^{\mu} k_{1}^{\nu} \ell \cdot (k_{1} + 4 k_{2}) \\
& \quad + \ell^{\mu} k_{2}^{\nu} \ell \cdot \left[ (2d - 4) k_{1} + (2d - 7) k_{2} \right] + k_{1}^{\mu} \ell^{\nu} \ell \cdot \left[ (2d - 7) k_{1} + 2 (d - 1) k_{2} \right] \\
& \quad + 4 k_{2}^{\mu} \ell^{\nu} \ell \cdot \left[ (d - 3) k_{1} + (d - 1) k_{2} \right] + \left[ k_{1}^{\mu} k_{1}^{\nu} + (d - 1) k_{1}^{\mu} k_{2}^{\nu} + 8 k_{2}^{\mu} k_{1}^{\nu} + (2d - 1) k_{2}^{\mu} k_{2}^{\nu} \right] \ell^{2} + \cdots \,.
}
Here we omit the linear and zeroth order terms, which do not give any divergence in the loop integral.
In ${\cal M}_{1}$, we also omit terms vanishing with $k_{1} \cdot \epsilon_{1} = k_{2} \cdot \epsilon_{2} = 0$ and $k_{1}^{2} = k_{2}^{2} = 0$ [including $(\ell \cdot \epsilon_{1}) (\ell \cdot k_{1})$, $(\ell \cdot k_{1})^{2}$, and so on].
It contributes to the divergent part as [see \cref{eq:loop-3,eq:loop-3-1,eq:loop-3-2}]
\eqs{
& h_{1} (x, y) = 5 (d - 1) \,, \\
& f_{1} (x, y) = 2 (6d - 1) (1- y) (1- y + xy) - 7 d (1- y + xy) - (5d - 2) (1- y) - 2 (3d - 8) \,, \\
& g_{1} (x, y) = (d + 4) (4d - 5) (1- y) (1- y + xy) - (3d - 8) (1- y + xy) - 4 (d - 1) (d + 3) (1- y) + 4 (3d - 5) \,.
}

\cref{fig:YM_loop} ($2$) gives
\eqs{
i {\cal M}_{2} =& i \epsilon (- i e^{2} {\tilde \mu}^{\epsilon}) \frac{1}{2} \int \frac{d^{d} \ell}{(2 \pi)^{d}} \frac{- i g_{\beta \sigma} \delta^{f d}}{\ell^{2}} \left[- \ell \cdot (\ell + p) g^{\alpha \beta} + \ell^{\alpha} (\ell + p)^{\beta} \right] \delta^{e f} \frac{- i g_{\alpha \rho} \delta^{e c}}{(\ell + p)^{2}} \epsilon_{1 \mu}^{*} \epsilon_{2 \nu}^{*} V_{e^{2}}^{(a \mu) (b \nu) (c \rho) (d \sigma)} \\
=& - \epsilon e^{2} T({\rm Ad}) \delta^{a b} {\tilde \mu}^{\epsilon}\int \frac{d^{d} \ell}{(2 \pi)^{d}} \frac{1}{[\ell^{2}] [(\ell + k_{2})^{2}] [(\ell + p)^{2}]} \left[ - (d - 2) g^{\mu \nu} (\ell^{2})^{2} - \ell^{\mu} \ell^{\nu} \ell^{2} \right. \\
& \left. - (d - 2) g^{\mu \nu} \ell^{2} \ell \cdot(k_{1} + 3 k_{2}) - 2 \ell^{\mu} \ell^{\nu} \ell \cdot k_{2} - \frac{1}{2} \ell^{\mu} k_{1}^{\nu} \ell^{2} - \frac{1}{2} k_{2}^{\mu} \ell^{\nu} \ell^{2} - 2 (d - 2) g^{\mu \nu} (\ell \cdot k_{1}) \ell \cdot k_{2}  \right. \\
& \left. - \ell^{\mu} k_{1}^{\nu} \ell \cdot k_{2} - k_{2}^{\mu} \ell^{\nu} \ell \cdot k_{2} + \cdots \right] \,.
}
In the first equality, we take into account the symmetric factor of $1/2$.
In the second equality, we multiply $(\ell + k_{2})^{2}$ in both the numerator and denominator of the integrand, and use
\eqs{
& g_{\beta \sigma} g_{\alpha \rho} \delta^{f d} \delta^{e f} \delta^{e c} \left[- \ell \cdot (\ell + p) g^{\alpha \beta} + \ell^{\alpha} (\ell + p)^{\beta} \right] V_{e^{2}}^{(a \mu) (b \nu) (c \rho) (d \sigma)} (\ell + k_{2})^{2} \\
& = T({\rm Ad}) \delta^{a b} \left[ - 2 (d - 2) g^{\mu \nu} \ell^{2} - 2 \ell^{\mu} \ell^{\nu}  - 2 (d - 2) g^{\mu \nu} \ell \cdot p - \ell^{\mu} p^{\nu} - p^{\mu} \ell^{\nu} \right] (\ell + k_{2})^{2} \\
& = T({\rm Ad}) \delta^{a b} \left[ - 2 (d - 2) g^{\mu \nu} (\ell^{2})^{2} - 2 \ell^{\mu} \ell^{\nu} \ell^{2} - 2 (d - 2) g^{\mu \nu} \ell^{2} \ell \cdot(k_{1} + 3 k_{2}) - 4 \ell^{\mu} \ell^{\nu} \ell \cdot k_{2} \right. \\
& \quad \left. - \ell^{\mu} (k_{1} + k_{2})^{\nu} \ell^{2} - (k_{1} + k_{2})^{\mu} \ell^{\nu} \ell^{2} - 4 (d - 2) g^{\mu \nu} \left[\ell \cdot (k_{1} + k_{2}) \right] \ell \cdot k_{2} - 2 (d - 2) g^{\mu \nu} \ell^{2} k_{2}^{2} \right. \\
& \quad \left. - \ell^{\mu} \ell^{\nu} k_{2}^{2} - 2 \ell^{\mu} (k_{1} + k_{2})^{\nu} \ell \cdot k_{2} - 2 (k_{1} + k_{2})^{\mu} \ell^{\nu} \ell \cdot k_{2} + \cdots \right]
\,.
}
Here we omit the linear-order terms, which do not give any divergence in the loop integral.
In ${\cal M}_{2}$, we also omit terms vanishing with $k_{1} \cdot \epsilon_{1} = k_{2} \cdot \epsilon_{2} = 0$ and $k_{1}^{2} = k_{2}^{2} = 0$.
It contributes to the divergent part as [see \cref{eq:loop-3,eq:loop-3-1,eq:loop-3-2}]
\eqs{
& h_{2} (x, y) = - (d - 1)^{2} \,, \\
& f_{2} (x, y) = - 2 (2 d^{2} - 7) (1- y) (1- y + xy) + (d - 2) (d + 2) (1- y + xy) + (3d^{2} - 10) (1- y) - 2 (d - 2) \,, \\
& g_{2} (x, y) = - (d + 4) (1- y) (1- y + xy) + \frac{d + 2}{2} (1- y + xy) + \frac{d + 6}{2} (1- y) - 1 \,.
}

\cref{fig:YM_loop} ($3$) and ($3'$) are from the insertion of $T^{\mu}_{~ \mu} \supset - \epsilon  {\tilde \mu}^{\epsilon / 2} e f^{a b c} A^{a \mu} A^{b \nu} \partial_{\mu} A^{c}_{\nu}$.
\cref{fig:YM_loop} (3) gives
\eqs{
i {\cal M}_{3} =& \epsilon e {\tilde \mu}^{\epsilon / 2} (e {\tilde \mu}^{\epsilon / 2}) \frac{1}{2} \int \frac{d^{d} \ell}{(2 \pi)^{d}} \frac{- i g_{\alpha \rho} \delta^{c e}}{\ell^{2}} \frac{- i g_{\beta \sigma} \delta^{d f}}{(\ell + k_{2})^{2}} \epsilon_{1 \mu}^{*}  f^{a c d} V_{e}^{\mu \alpha \beta} (- k_{1}, \ell, - \ell - k_{2}) \epsilon_{2 \nu}^{*} f^{b e f}  V_{e}^{\nu \rho \sigma} (- k_{2}, - \ell, \ell + k_{2}) \\
=& - \epsilon e^{2} T({\rm Ad}) \delta^{a b} {\tilde \mu}^{\epsilon}\int \frac{d^{d} \ell}{(2 \pi)^{d}} \frac{1}{[\ell^{2}] [(\ell + k_{2})^{2}] [(\ell + p)^{2}]} \left[ - g^{\mu \nu} (\ell^{2})^{2} - (2d - 3) \ell^{\mu} \ell^{\nu} \ell^{2} \right. \\
& \left. - g^{\mu \nu} \ell^{2} \ell \cdot (2 k_{1} + 3 k_{2}) - 2 (2d - 3) \ell^{\mu} \ell^{\nu} \ell \cdot (k_{1} + k_{2}) - \frac{2d - 3}{2} k_{2}^{\mu} \ell^{\nu} \ell^{2} - \frac{1}{2} g^{\mu \nu} \ell^{2} k_{1} \cdot k_{2} \right. \\
& \left. - 2 g^{\mu \nu} (\ell \cdot k_{1}) \ell \cdot k_{2} - 2 (2d - 3) \ell^{\mu} \ell^{\nu} k_{1} \cdot k_{2} - (2d - 3) k_{2}^{\mu} \ell^{\nu} \ell \cdot k_{1} - \frac{3}{2} k_{2}^{\mu} k_{1}^{\nu} \ell^{2} + \cdots \right] \,.
}
In the first equality, we take into account the symmetric factor of $1/2$.
In the second equality, we multiply $(\ell + p_{2})^{2}$ in both the numerator and denominator of the integrand, and use $ \delta^{c e} \delta^{d f} f^{a c d} f^{b e f} = T({\rm Ad}) \delta^{a b}$ and
\eqs{
& g_{\alpha \rho}  g_{\beta \sigma} V_{e}^{\mu \alpha \beta} (- k_{1}, \ell, - \ell - k_{2}) V_{e}^{\nu \rho \sigma} (- k_{2}, - \ell, \ell + k_{2}) (\ell + p)^{2} \\
& = \left[ - 2 g^{\mu \nu} \ell^{2} - 2 (2d - 3) \ell^{\mu} \ell^{\nu} - 2 g^{\mu \nu} \ell \cdot k_{2} - (2d - 3) \ell^{\mu} k_{2}^{\nu} - (2d - 3) k_{2}^{\mu} \ell^{\nu} \right. \\
& \quad \left. + g^{\mu \nu} (3 k_{1} \cdot k_{2} - 2 k_{2}^{2})  - 3 k_{2}^{\mu}  k_{1}^{\nu} - (d - 3) k_{2}^{\mu} k_{2}^{\nu} \right] (\ell + k_{1} + k_{2})^{2} \\
& = - 2 g^{\mu \nu} (\ell^{2})^{2} - 2 (2d - 3) \ell^{\mu} \ell^{\nu} \ell^{2} - 2 g^{\mu \nu} \ell^{2} \ell \cdot (2 k_{1} + 3 k_{2}) - 4 (2d - 3) \ell^{\mu} \ell^{\nu} \ell \cdot (k_{1} + k_{2}) \\
& \quad - (2d - 3) \ell^{\mu} k_{2}^{\nu} \ell^{2} - (2d - 3) k_{2}^{\mu} \ell^{\nu} \ell^{2} - g^{\mu \nu} \ell^{2} (2 k_{1}^{2} + k_{1} \cdot k_{2} + 4 k_{2}^{2}) - 4 g^{\mu \nu} \left[\ell \cdot (k_{1} + k_{2}) \right] \ell \cdot k_{2} \\
& \quad - 2 (2d - 3) \ell^{\mu} \ell^{\nu} (k_{1} + k_{2})^{2}  - 2 (2d - 3) \ell^{\mu} k_{2}^{\nu} \ell \cdot (k_{1} + k_{2}) - 2 (2d - 3) k_{2}^{\mu} \ell^{\nu} \ell \cdot (k_{1} + k_{2}) \\
& \quad  - 3 k_{2}^{\mu} k_{1}^{\nu} \ell^{2} - (d - 3) k_{2}^{\mu} k_{2}^{\nu} \ell^{2} + \cdots \,.
}
Here we omit the linear and zeroth-order terms, which do not give any divergence in the loop integral.
In ${\cal M}_{3}$, we also omit terms vanishing with $k_{1} \cdot \epsilon_{1} = k_{2} \cdot \epsilon_{2} = 0$ and $k_{1}^{2} = k_{2}^{2} = 0$.
It contributes to the divergent part as [see \cref{eq:loop-3,eq:loop-3-1,eq:loop-3-2}]
\eqs{
h_{3} (x, y) & = - 3 (d - 1) \,, \\
f_{3} (x, y) & = - 2 (4d + 1) (1- y) (1- y + xy) + 2 (3d - 1) (1- y + xy) + 7d (1- y) - \frac{9d - 8}{2} \,, \\
g_{3} (x, y) & = - (d + 4) (2d - 3) (1- y) (1- y + xy) + 2 (2d - 3) (1- y + xy) \\
& \qquad + \frac{(d + 6) (2d - 3)}{2} (1- y) - \frac{7d - 6}{2} \,.
}

\cref{fig:YM_loop} ($3'$) gives
\eqs{
i {\cal M}_{3'} =& \epsilon e {\tilde \mu}^{\epsilon / 2} (e {\tilde \mu}^{\epsilon / 2}) \frac{1}{2} \int \frac{d^{d} \ell}{(2 \pi)^{d}} \frac{- i g_{\alpha \rho} \delta^{c e}}{(\ell + p)^{2}} \frac{- i g_{\beta \sigma} \delta^{d f}}{(\ell + k_{2})^{2}} \epsilon_{1 \mu}^{*}  f^{a c d} V_{e}^{\mu \alpha \beta} (- k_{1}, \ell + p, - \ell - k_{2}) \\
& \times \epsilon_{2 \nu}^{*} f^{b e f}  V_{e}^{\nu \rho \sigma} (- k_{2}, - \ell - p, \ell + k_{2}) \\
=& - \epsilon e^{2} T({\rm Ad}) \delta^{a b} {\tilde \mu}^{\epsilon}\int \frac{d^{d} \ell}{(2 \pi)^{d}} \frac{1}{[\ell^{2}] [(\ell + k_{2})^{2}] [(\ell + p)^{2}]} \left[ - g^{\mu \nu} (\ell^{2})^{2} - (2d - 3) \ell^{\mu} \ell^{\nu}\ell^{2} \right. \\
& \left. - g^{\mu \nu} \ell^{2} \ell \cdot (k_{1} + 2 k_{2}) - \frac{2d - 3}{2} \ell^{\mu} k_{1}^{\nu} \ell^{2} - (2d - 3) k_{2}^{\mu} \ell^{\nu} \ell^{2} + \frac{1}{2} g^{\mu \nu} \ell^{2} k_{1} \cdot k_{2} - d \ell^{2} k_{2}^{\mu} k_{1}^{\nu} \right] \,.
}
In the first equality, we take into account the symmetric factor of $1/2$.
In the second equality, we multiply $\ell^{2}$ in both the numerator and denominator of the integrand, and use $ \delta^{c e} \delta^{d f} f^{a c d} f^{b e f} = T({\rm Ad}) \delta^{a b}$ and
\eqs{
& g_{\alpha \rho}  g_{\beta \sigma} V_{e}^{\mu \alpha \beta} (- k_{1}, \ell + p, - \ell - k_{2}) V_{e}^{\nu \rho \sigma} (- k_{2}, - \ell - p, \ell + k_{2}) \\
& = \left[ - 2 g^{\mu \nu} \ell^{2} - 2 (2d - 3) \ell^{\mu} \ell^{\nu} - 2 g^{\mu \nu} \ell \cdot (k_{1} + 2 k_{2}) - (2d - 3) \ell^{\mu} (k_{1} + 2 k_{2})^{\nu} - (2d - 3) (k_{1} + 2 k_{2})^{\mu} \ell^{\nu} \right. \\
& \quad \left. - g^{\mu \nu} (2 k_{1}^{2} - k_{1} \cdot k_{2} + 2 k_{2}^{2}) - (d - 3) k_{1}^{\mu} k_{1}^{\nu} - (2d - 3) k_{1}^{\mu} k_{2}^{\nu} - 2d k_{2}^{\mu} k_{1}^{\nu} - 2 (2d - 3) k_{2}^{\mu} k_{2}^{\nu} \right] \,.
}
In ${\cal M}_{3'}$, we also omit terms vanishing with $k_{1} \cdot \epsilon_{1} = k_{2} \cdot \epsilon_{2} = 0$ and $k_{1}^{2} = k_{2}^{2} = 0$.
It contributes to the divergent part as [see \cref{eq:loop-3,eq:loop-3-1,eq:loop-3-2}]
\eqs{
h_{3'} (x, y) & = - 3 (d - 1) \,, \\
f_{3'} (x, y) & = - 2 (4d + 1) (1- y) (1- y + xy) + (d + 2) (1- y + xy) + 2 (d + 2) (1- y) + \frac{d}{2} \,, \\
g_{3'} (x, y) & = - (d + 4) (2d - 3) (1- y) (1- y + xy) + \frac{(d + 2) (2d - 3)}{2} (1- y + xy) \\
& \qquad + (d + 2) (2d - 3) (1- y) - d^{2} \,.
}

\cref{fig:YM_loop} ($4$) is from the insertion of $T^{\mu}_{~ \mu} \supset - (1/4) \epsilon  {\tilde \mu}^{\epsilon} e^{2}  f^{a b e} f^{c d e} A^{a}_{\mu} A^{b}_{\nu} A^{c \mu} A^{d \nu}$.
\cref{fig:YM_loop} ($4$) gives
\eqs{
i {\cal M}_{4} =& - i \epsilon e^{2} {\tilde \mu}^{\epsilon} \frac{1}{2} \int \frac{d^{d} \ell}{(2 \pi)^{d}} \frac{- i g_{\rho \sigma} \delta^{c d}}{\ell^{2}} V_{e^{2}}^{(a \mu) (b \nu) (c \rho) (d \sigma)} \\
=& - \epsilon e^{2} T({\rm Ad}) \delta^{a b} {\tilde \mu}^{\epsilon}\int \frac{d^{d} \ell}{(2 \pi)^{d}} \frac{1}{[\ell^{2}] [(\ell + k_{2})^{2}] [(\ell + p)^{2}]} \left[ (d - 1) g^{\mu \nu} (\ell^{2})^{2} + 2 (d - 1) g^{\mu \nu} \ell^{2} \ell \cdot (k_{1} + 2 k_{2}) \right. \\
& \left. + 2 (d - 1) g^{\mu \nu} \ell^{2} k_{1} \cdot k_{2} + 4 (d - 1) g^{\mu \nu} (\ell \cdot k_{1}) \ell \cdot k_{2} + \cdots \right] \,.
}
In the first equality, we take into account the symmetric factor of $1/2$.
In the second equality, we multiply $(\ell + k_{2})^{2} (\ell + p)^{2}$ in both the numerator and denominator of the integrand, and use 
\eqs{
& g_{\rho \sigma} \delta^{c d} V_{e^{2}}^{(a \mu) (b \nu) (c \rho) (d \sigma)} \\
& = 2 (d - 1) T({\rm Ad}) \delta^{a b} g^{\mu \nu} (\ell + k_{2})^{2} (\ell + k_{1} + k_{2})^{2} \\
& = 2 (d - 1) T({\rm Ad}) \delta^{a b} g^{\mu \nu} \left[ (\ell^{2})^{2} + 2 \ell^{2} \ell \cdot (k_{1} + 2 k_{2}) + \ell^{2} (k_{1}^{2} + 2 k_{1} \cdot k_{2} + k_{2}^{2}) + 4 \left[ \ell \cdot (k_{1} + k_{2})\right] \ell \cdot k_{2} + \cdots \right] \,.
}
Here we omit the linear and zeroth-order terms, which do not give any divergence in the loop integral.
In ${\cal M}_{4}$, we also omit terms vanishing with $k_{1} \cdot \epsilon_{1} = k_{2} \cdot \epsilon_{2} = 0$ and $k_{1}^{2} = k_{2}^{2} = 0$.
It contributes to the divergent part as [see \cref{eq:loop-3,eq:loop-3-1,eq:loop-3-2}]
\eqs{
& h_{4} (x, y) = d (d - 1) \,, \\
& f_{4} (x, y) = 4 (d - 1) (d + 2) (1- y) (1- y + xy) - 2 (d - 1) (d + 2) (1- y + xy) - 4 (d - 1)(d + 2) (1- y) \\ 
& \quad \quad \quad \quad ~ + 2 (d - 1) (d + 2) \\
& g_{4} (x, y) = 0 \,.
}

In summary,
\eqs{
& h (x, y) = 0 \quad \text{(i.e., no quadratic divergence),} \\
& f (x, y) = - d (d + 2) (1- y + xy) - (d - 2) (d + 2) (1- y) + 2 (d^{2} - 5 d + 10) \,, \\
& g(x, y) = d (d + 2) (1- y + xy) - d (d + 2) (1- y) - \frac{2 d^{2} - 17 d + 36}{2} \,.
}
We obtain the divergent part of
\eqs{
\left. i {\cal M}_{\rm loop} \right|_{\rm div} =& - \epsilon \frac{1}{2} i \frac{e^{2}}{16 \pi^{2}} e^{\gamma_{E} (\epsilon / 2)} \Gamma(\epsilon / 2)T({\rm Ad}) \delta^{a b} \epsilon_{1 \mu}^{*} \epsilon_{2 \nu}^{*} \int_{0}^{1} dx dy y \, \left[ f (x, y) k_{1} \cdot k_{2} g^{\mu \nu} + g (x, y) k_{2}^{\mu} k_{1}^{\nu} \right] \left( \frac{\mu^{2}}{\Delta_{xy}} \right)^{\epsilon / 2} \,,
} where 
\eqs{
& f (x, y) = (2 - x) y (d - 4)^{2} - 2 (6 - 9 y + 5 xy) (d - 4) - 12 (2 - 3 y + 2 xy) \,, \\
& g(x, y) = - (1 - x y) (d - 4)^{2} + \frac{1+ 20 xy}{2} (d - 4) + 24 xy  \,, \\
& \int_{0}^{1} dx dy \, y \, f (x, y) = \frac{1}{2} (d - 4)^{2} - \frac{5}{3} (d - 4) - 4 \,, \\
& \int_{0}^{1} dx dy \, y \, g (x, y) = - \frac{1}{3} (d - 4)^{2} + \frac{23}{12} (d - 4) + 4 \,,
}
and thus,
\eqs{
i {\cal M}_{\rm loop} =& i \frac{e^{2}}{16 \pi^{2}} 4 T({\rm Ad}) (k_{1} \cdot k_{2} g^{\mu \nu} - k_{2}^{\mu} k_{1}^{\nu}) \epsilon_{1 \mu}^{*} \epsilon_{2 \nu}^{*}  \delta^{a b} \,.
}
Combining the type-1) [see \cref{eq:YM_tree}] and 2) contributions, we find
\eqs{
i {\cal M}_{\rm tree} + i {\cal M}_{\rm loop} = i \left( \frac{22}{3} T({\rm Ad}) - \frac{4}{3} T(F) - \frac{2}{3} T(S) \right) \frac{e^{2}}{16 \pi^{2}} (k_{1} \cdot k_{2} g^{\mu \nu} - k_{2}^{\mu} k_{1}^{\nu}) \epsilon_{1 \mu}^{*} \epsilon_{2 \nu}^{*}  \delta^{a b} \,,
}
which is reproduced by the insertion of $(\beta_{e} / 2e) F_{\mu \nu}^{a 2}$ [see \cref{eq:betae_YM}] as expected.

\subsection{Divergent one-loop integrals \label{sec:loopfunc_YM}}

The two-point integrals are
\eqs{
\label{eq:loop-2}
\left. {\tilde \mu}^{\epsilon} \int \frac{d^{d} \ell}{(2 \pi)^{d}} \frac{1}{[\ell^{2}] [(\ell + p)^{2}]} \right|_{\rm div} =& \frac{i}{16 \pi^{2}} e^{\gamma_{E} (\epsilon / 2)} \Gamma(\epsilon / 2) \int_{0}^{1} dx \left( \frac{\mu^{2}}{\Delta_{x}} \right)^{\epsilon / 2} \,, \\
\left. {\tilde \mu}^{\epsilon} \int \frac{d^{d} \ell}{(2 \pi)^{d}} \frac{\ell^{\mu}}{[\ell^{2}] [(\ell + p)^{2}]} \right|_{\rm div} =& \frac{i}{16 \pi^{2}} e^{\gamma_{E} (\epsilon / 2)} \Gamma(\epsilon / 2) \int_{0}^{1} dx \, (- q_{x}^{\mu}) \left( \frac{\mu^{2}}{\Delta_{x}} \right)^{\epsilon / 2} \,, \\
\left. {\tilde \mu}^{\epsilon} \int \frac{d^{d} \ell}{(2 \pi)^{d}} \frac{\ell^{\mu} \ell^{\nu}}{[\ell^{2}] [(\ell + p)^{2}]} \right|_{\rm div} =& \frac{i}{16 \pi^{2}} e^{\gamma_{E} (\epsilon / 2)} \left( - \frac{1}{2} \right) \Gamma(- 1 +  \epsilon / 2) \int_{0}^{1} dx \, g^{\mu \nu} \Delta_{x} \left( \frac{\mu^{2}}{\Delta_{x}} \right)^{\epsilon / 2} \\
& + \frac{i}{16 \pi^{2}} e^{\gamma_{E} (\epsilon / 2)} \Gamma(\epsilon / 2) \int_{0}^{1} dx \, q_{x}^{\mu} q_{x}^{\nu} \left( \frac{\mu^{2}}{\Delta_{x}} \right)^{\epsilon / 2} \,.
}
where $q_{x} = x p$ and $\Delta_{x} = - x (1 -x) p^{2}$.

We use%
\footnote{
The version with $x \to x / (1 - y)$ and $y \to (1 - y)$, i.e., $\int_{0}^{1} dx dy dz \delta(1 - x - y - z) \dots$ may also be familiar to readers.
}
\eqs{
\frac{1}{A B C} = \int_{0}^{1} dx dy \, 2y \frac{1}{[(1 - x) y A + xy B + (1 - y) C]^{3}} \,.
}
The three-point integrals are
\eqs{
\label{eq:loop-3}
\left. {\tilde \mu}^{\epsilon} \int \frac{d^{d} \ell}{(2 \pi)^{d}} \frac{1; \ell^{\mu}}{[\ell^{2}] [(\ell + k_{2})^{2}] [(\ell + p)^{2}]} \right|_{\rm div} =& 0 \,, \\
\left. {\tilde \mu}^{\epsilon} \int \frac{d^{d} \ell}{(2 \pi)^{d}} \frac{\ell^{\mu} \ell^{\nu}}{[\ell^{2}] [(\ell + k_{2})^{2}] [(\ell + p)^{2}]} \right|_{\rm div} =& \frac{1}{2} \frac{i}{16 \pi^{2}} e^{\gamma_{E} (\epsilon / 2)} \Gamma(\epsilon / 2) \int_{0}^{1} dx dy y \, g^{\mu \nu}  \left( \frac{\mu^{2}}{\Delta_{xy}} \right)^{\epsilon / 2} \,, \\
\left. {\tilde \mu}^{\epsilon} \int \frac{d^{d} \ell}{(2 \pi)^{d}} \frac{\ell^{\mu} \ell^{\nu} \ell^{\rho}}{[\ell^{2}] [(\ell + k_{2})^{2}] [(\ell + p)^{2}]} \right|_{\rm div} =& \frac{1}{2} \frac{i}{16 \pi^{2}} e^{\gamma_{E} (\epsilon / 2)} \Gamma(\epsilon / 2) \int_{0}^{1} dx dy y \, \\
& \times - \left( q_{xy}^{\mu} g^{\nu \rho} + q_{xy}^{\nu} g^{\mu \rho} + q_{xy}^{\rho} g^{\mu \nu} \right)\left( \frac{\mu^{2}}{\Delta_{xy}} \right)^{\epsilon / 2} \,, \\
\left. {\tilde \mu}^{\epsilon} \int \frac{d^{d} \ell}{(2 \pi)^{d}} \frac{\ell^{2} \ell^{\mu} \ell^{\nu}}{[\ell^{2}] [(\ell + k_{2})^{2}] [(\ell + p)^{2}]} \right|_{\rm div} =& \frac{i}{16 \pi^{2}} e^{\gamma_{E} (\epsilon / 2)} \left( - \frac{d + 2}{4} \right) \Gamma(- 1 +  \epsilon / 2) \int_{0}^{1} dx dy y \, g^{\mu \nu} \Delta_{xy} \left( \frac{\mu^{2}}{\Delta_{xy}} \right)^{\epsilon / 2} \\
& + \frac{1}{2} \frac{i}{16 \pi^{2}} e^{\gamma_{E} (\epsilon / 2)} \Gamma(\epsilon / 2) \int_{0}^{1} dx dy y \, \left[q_{xy}^{2} g^{\mu \nu} + (d + 4) q_{xy}^{\mu} q_{xy}^{\nu} \right] \left( \frac{\mu^{2}}{\Delta_{xy}} \right)^{\epsilon / 2} \,.
}
where $p = k_{1} + k_{2}$, $q_{xy} = (1 - y) k_{1} + (1 - y + xy) k_{2}$, and $\Delta_{xy} = - y (1 - x) (1 - y) p^{2} - x (1 - x) y^{2} k_{1}^{2} - x y (1- y) k_{2}^{2}$.

Here are useful expressions when determining $f (x, y)$, $g (x, y)$, and $h (x, y)$.
For $k_{1}^{2} = k_{2}^{2} = 0$,
\eqs{
\label{eq:loop-3-1}
& q_{xy}^{2} = 2 (1- y) (1- y + xy) k_{1} \cdot k_{2} \,, \quad q_{xy} \cdot k_{1} = (1- y + xy) k_{1} \cdot k_{2} \,, \quad q_{xy} \cdot k_{2} = (1- y) k_{1} \cdot k_{2} \,.
}
For the contraction of $\epsilon_{1 \mu} \epsilon_{2 \nu}$ and $k_{1} \cdot \epsilon_{1} = k_{2} \cdot \epsilon_{2} = 0$:
\eqs{
\label{eq:loop-3-2}
\quad q_{xy}^{\mu} q_{xy}^{\nu} = (1- y) (1- y + xy) k_{2}^{\mu} k_{1}^{\nu} \,, \quad q_{xy}^{\mu} k_{1}^{\nu} = (1- y + xy) k_{2}^{\mu} k_{1}^{\nu} \,, \quad k_{2}^{\mu} q_{xy}^{\nu} = (1- y) k_{2}^{\mu} k_{1}^{\nu} \,,
 }
and $k_{1}^{\mu} q_{xy}^{\nu} = q_{xy}^{\mu} k_{2}^{\nu} = 0$.

We use%
\footnote{
The version with $x \to x / (1 - y - z)$, $y \to (1 - y - z) / (1 - z)$, and $z \to 1 - z$, i.e., $\int_{0}^{1} dx dy dz dw \delta(1 - x - y - z - w) \dots$ may also be familiar to readers.
}
\eqs{
\frac{1}{A B C D} = \int_{0}^{1} dx dy dz \, 6yz^{2} \frac{1}{[(1 - x) y z A + xyz B + (1 - z) C + (1 - y)z D ]^{4}} \,.
}
The four-point integral is
\eqs{
\left. {\tilde \mu}^{\epsilon} \int \frac{d^{d} \ell}{(2 \pi)^{d}} \frac{\ell^{2} \ell^{\mu} \ell^{\nu}}{[\ell^{2}] [(\ell + k)^{2}] [(\ell + p)^{2}] [(\ell + q)^{2}]} \right|_{\rm div} =& \frac{d + 2}{4} \frac{i}{16 \pi^{2}} e^{\gamma_{E} (\epsilon / 2)} \Gamma(\epsilon / 2) \int_{0}^{1} dx dy dz yz^{2} g^{\mu \nu}\left( \frac{\mu^{2}}{\Delta_{xyz}} \right)^{\epsilon / 2} \,,
}
where $\Delta_{xyz} = - xyz (1 - xyz) k^{2} - z (1 - z) p^{2} - (1 - y) z (1 + yz - z) q^{2} + 2 xyz (1 - z) k \cdot p + 2 xy (1 - y) z^{2} k \cdot q + 2 (1 - y) z (1 - z) p \cdot q$.

\end{document}